\documentclass[journal=jacsat,manuscript=article,layout=onecolumn]{achemso}
\usepackage{color}

\usepackage[utf8]{inputenc}
\usepackage{graphicx}
\usepackage{xcolor}

\usepackage[version=3]{mhchem} 
\usepackage[T1]{fontenc}       
\usepackage{graphicx}
\usepackage{subfigure}
\usepackage{multirow}
\usepackage{hhline}
\usepackage{chemarr}
\usepackage{amssymb}
\usepackage{color}
\usepackage{xspace}
\usepackage{chngcntr}
\counterwithout{table}{section}
\usepackage{textcomp}
\usepackage{gensymb}
\usepackage{xr}
\usepackage[normalem]{ulem}

\usepackage{caption}
\usepackage{booktabs}

\title{Water Dynamics around T0 vs. R4 of Hemoglobin from Local
  Hydrophobicity Analysis} \author{Seyedeh Maryam Salehi, Marco
  Pezzella, Adam Willard, Markus Meuwly}\email{m.meuwly@unibas.ch}
\author{Martin Karplus}\email{marci@tammy.harvard.edu}

\begin{document}

\date{\today}

\maketitle

\begin{abstract}
The local hydration around tetrameric Hb in its T$_0$ and R$_4$
conformational substates is analyzed based on molecular dynamics
simulations. Analysis of the local hydrophobicity (LH) for all
residues at the $\alpha_1 \beta_2$ and $\alpha_2 \beta_1$ interfaces,
responsible for the quaternary T$\rightarrow$R transition, which is
encoded in the MWC model, as well as comparison with earlier
computations of the solvent accessible surface area (SASA), makes
clear that the two quantities measure different aspects of
hydration. Local hydrophobicity quantifies the presence and structure
of water molecules at the interface whereas ``buried surface'' reports
on the available space for solvent. For simulations with Hb frozen in
its T$_0$ and R$_4$ states the correlation coefficient between LH and
buried surface is 0.36 and 0.44, respectively, but it increases
considerably if the 95 \% confidence interval is used. The LH with Hb
frozen and flexible changes little for most residues at the interfaces
but is significantly altered for a few select ones, which are
Thr41$\alpha$, Tyr42$\alpha$, Tyr140$\alpha$, Trp37$\beta$,
Glu101$\beta$ (for T$_0$) and Thr38$\alpha$, Tyr42$\alpha$,
Tyr140$\alpha$ (for R$_4$). The number of water molecules at the
interface is found to increase by $\sim 25$ \% for
T$_0$$\rightarrow$R$_4$ which is consistent with earlier
measurements. Since hydration is found to be essential to protein
function, it is clear that hydration also plays an essential role in
allostery.
\end{abstract}

\section{Introduction}
Hydration is important for protein function. It has been reported that
at least one monolayer of water is required for a protein to
function.\cite{teeter:1991} The properties of solvent water near the
protein surface have been characterized experimentally - by nuclear
magnetic resonance (NMR), quasi inelastic neutron
scattering\cite{caronna:2005} and M\"ossbauer
spectroscopy\cite{parak:2002} and computationally with molecular
dynamics (MD)
simulations.\cite{mouawad:2005,karplus.hb:2011,yusuff:2012,Hub_2010,MM.hb:2018,MM.hb:2020}
In the crowded cellular environment, the average separation of
macromolecules is of the order of 10 \AA\/, which corresponds to only
$\sim 3$ layers of water molecules. From the NMR experiments and MD
simulations it was found that the reorientation dynamics of water on
the protein surface is slowed down by a factor of 2 to 3 compared with
water in the bulk. It is notable that although it has been known for
almost 60 years\cite{bernal:1965} that the dynamics of water adjacent
to a macromolecule differs from that in the bulk, as of now only
little is known about the special properties of cellular
water.\cite{stadler:2008}\\

\noindent
Hemoglobin (Hb), which is physiologically involved in oxygen (O$_2$)
transport, is a widely studied protein for which a broad range of
molecularly resolved studies are available. The tetramer consists of
two $(\alpha \beta)$ homodimers $(\alpha_1 \beta_1)$ and $(\alpha_2
\beta_2)$ which are referred to as ``subunit 1'' (S1) and ``subunit
2'' (S2) in the following. Functionally most relevant are the two
endpoint structures T$_0$ and R$_4$ which correspond to the
ligand-free and fully ligand-bound proteins, respectively. The
monomer-monomer interfaces ($\alpha_1 \beta_1$ and $\alpha_2 \beta_2$)
do not change during the T$_0$$\rightarrow$R$_4$ transition whereas
the dimer-dimer interface changes appreciably due to what can be
described as a $15^\circ$ rotation of S1 versus S2, although the
actual transition is more complicated.\cite{karplus.hb:2011} The
quaternary structural transition is accompanied by a change in
exposure to hydration of residues lining the protein/solvent interface
and by a change in the solvent accessible surface area. This change in
solvent exposure is thought to contribute to the difference in
thermodynamic stability of the two conformational substates T$_0$ and
R$_4$.\cite{chothia:1985}\\

\noindent
The change in solvent exposure is also of interest for the two
unligated forms T$_0$ and R$_0$. Experimentally, the T$_0$ state was
found to be more stable than R$_0$ by $\sim 7$ kcal/mol when 2,3-DPG
is bound to the tetramers,\cite{edelstein:1971} which is reduced to
$\sim 3.5$ kcal/mol without 2,3-DPG bound to HbA\cite{kister:1987}.
This is in striking disagreement with a number of all-atom MD
simulations that reported unstable T$_0$ structures on the hundreds of
ns time scale.\cite{Hub_2010,yusuff:2012} The role of solvent in
stabilizing one conformational substate over another one was already
noted about 50 years ago:\cite{janin:1976} ``A larger surface area is
buried in deoxy- than in methemoglobin as a result of tertiary and
quaternary structure changes. [..] This implies that hydrophobicity
stabilizes the deoxy structure, the free energy spent in keeping the
subunits in a low-affinity conformation being compensated by
hydrophobic free energy due to the smaller surface area accessible to
solvent.'' In other words, the ``hydrophobic
effect'',\cite{janin:1976,chothia:1985} which arises from the
disruption of the bulk water hydrogen bond network around nonpolar
groups,\cite{Rossky:1979,rossky:1998} is likely to be a major driving
force underlying differential stabilization of T$_0$ over R$_0$ and
R$_4$. The theoretical analysis of Chandler and
coworkers\cite{chandler:2005,chandler.varenna:2012} indicated that for
large molecules, there was a "dewetting" phenomenon that stabilizes
compact (T-state) relative to more open (R-state) structures.\\

\noindent
Since hydration is required for a protein to function, it is clear
that hydration is essential for the allosteric transition from T$_0$
to R$_4$ to occur. The active role water plays in biological processes
has been discussed previously for protein-ligand binding, in
particular. With its hydrogen bond-donor and acceptor capabilities,
individual water molecules are highly adaptable at interfaces. It has
been found that water can act as an extension to the protein
structure.\cite{ladbury:1996} At the host / water interface pronounced
density fluctuations can occur which manifest themselves in
time-varying occupational and orientational water dynamics. More
recently, MD simulations together with machine learning analyses have
been combined for a deeper understanding of water molecules at
protein-ligand interfaces.\cite{lill:2020,rizzi:2021} As an example,
six ligands with an octa-acid calixarene host have been considered and
it was found that the relevant collective variables describing the
ligand-bound and the ligand-free state differ.\cite{rizzi:2021} For
the unbound state the solvation around the ligand to enter together
with the number of water molecules in the cavity had a large weight in
the machine-learned model. Conversely, for the bound state the number
of water molecules around the cavity entrance are more
important. These findings indicate that it is valuable to analyse
explicit water motion near biological interfaces for a better
understanding of biological function.\\

\section{Results}
\noindent
The present work reports on the local hydrophobicity (LH) around Hb
from simulations of the $\alpha_1 \beta_1$ dimer and $\alpha_1 \beta_1
\alpha_2 \beta_2$ tetramer of the T$_0$ and R$_4$ structures. The main
questions quantified more precisely than in our earlier
study\cite{MM.hb:2020} concern a) the comparison of the local
hydrophobicities for rigid T$_0$ and R$_4$ in the MD simulations and
its relation to the analysis of the solvent accessible surface area
(SASA) by Lesk et al.;\cite{chothia:1985} and b) changes in LH that
arise when the proteins are flexible in the MD simulations; and c) the
changes in LH between isolated dimers S1 and S2 compared with those
for the tetramers in the two conformational substates.\\

\begin{table}[]
\caption{Position of interfacial residues inside the protein. Residues
  at the $\alpha_1 \beta_2$ / $\alpha_2 \beta_1$ interface are
  indicated with checks in the last column.}
\centering
\begin{tabular}{c||c|c|}
\hline
Residue     &  Position &  Positioned at \\
                  &                &  $\alpha_1 \beta_2$ or $\alpha_2 \beta_1$ Interface\\
\hline
Val1$\alpha$   &  N-terminus & \\
Pro37$\alpha$  &  3$_{10}$-helix & \checkmark  \\
Thr38$\alpha$ &  3$_{10}$-helix & \checkmark\\
Lys40$\alpha$  &  3$_{10}$-helix &\checkmark\\
Thr41$\alpha$  &  3$_{10}$-helix &\checkmark\\
Tyr42$\alpha$  &  3$_{10}$-helix &\checkmark\\
Pro44$\alpha$  &  Turn &\checkmark\\
Thr134$\alpha$ &  $\alpha$-helix & \\
Tyr140$\alpha$ &  Turn &\\
Arg141$\alpha$ &  C-Terminus &\\
\hline
Val1$\beta$   & N-terminus & \\
Trp37$\beta$  & 3$_{10}$-helix &\checkmark\\
Pro100$\beta$ & $\alpha$-helix &\checkmark\\
Glu101$\beta$ & $\alpha$-helix &\checkmark\\
Asn139$\beta$ & $\alpha$-helix & \\
Tyr145$\beta$ & Turn  &\checkmark\\
\hline
\end{tabular}
\label{tab:tabres}
\end{table}

\begin{figure}[H]
\begin{center}
\includegraphics[width=0.8\textwidth]{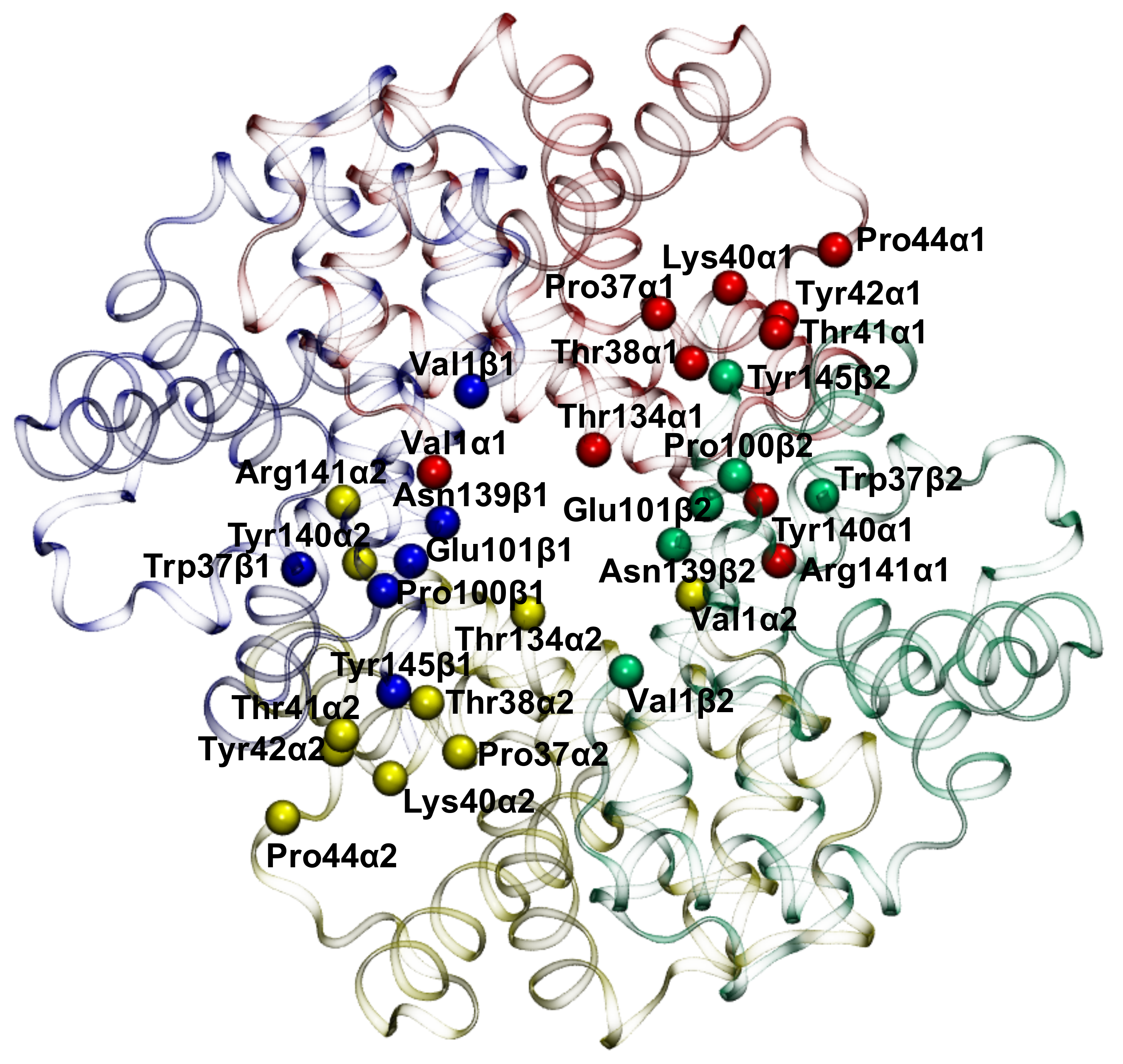}
\caption{Representation of T$_0$ Hb tetramer with the C$_{\alpha}$
  atoms of the studied residues as VDW spheres together with residue
  names. Red, blue, yellow and green ribbons and VDW spheres represent
  the $(\alpha_1 \beta_1)$ and $(\alpha_2 \beta_2)$ subunits S1 and S2
  of Hb.}
\label{fig:Hb}
\end{center}
\end{figure}

\noindent
In accord with the analysis of Lesk et al.\cite{chothia:1985} there
are 10 residues (Val1, Pro37, Thr38, Lys40, Thr41, Tyr42, Pro44,
Thr134, Tyr140, Arg141) that change significantly in solvent exposure
from buried to exposed in the $\alpha$ subunit interface in the
transition between T and R states, and 6 residues (Val1, Trp37,
Pro100, Glu101, Asn139, Tyr145) in the $\beta$ subunit
interface\cite{chothia:1985}; see Figure \ref{fig:Hb} for the
structure and labelled residues. For these residues a) the buried
surface as per the analysis in the literature\cite{chothia:1985} is
larger than 10 \AA\/$^2$ and b) the difference between the buried
surface for a given residue between the oxy and the deoxy structure
was found to be\cite{chothia:1985} larger than 20 \AA\/$^{2}$. These
criteria were used to select residues for analysis because for the
present work the {\it change in exposure} between the two
conformational substates is of interest. The positions in the protein
of the residues at the $\alpha_1 \beta_2$ and $\alpha_2 \beta_1$
interfaces are indicated in Table \ref{tab:tabres}.  Additional
residues that are of potential interest but were not included in the
analysis are Asp126$\alpha$, Lys139$\alpha$, Lys82$\beta$,
Tyr145$\beta$, and His146$\beta$.\\

\noindent
Water dynamics, which can be obtained from MD simulations, is used to
quantitatively determine the role of hydrophobicity in the T$_0$ and
R$_4$ states of Hb. For this purpose, the time-resolved displacements
of water molecules at the protein-solvent interface and the coupling
of these displacements with rearrangements in the protein subunits are
investigated.\\

\noindent
Previously, the solvent exposure of buried and exposed interfacial
residues for T$_0$ and R$_4$ was analyzed by computing the solvent
accessible surface area (SASA) for the available X-ray structures and
was reported to correlate with protein stability.\cite{chothia:1985}
To probe water dynamics for the native state of the protein, and to
estimate the local hydrophobicity without the influence of the protein
conformational degrees of freedom, simulations in which the protein
degrees were fixed (``frozen'') were performed. Simulations in which
the protein degrees of freedom were not held fixed (``flexible') were
also carried out; they include entropic contributions to local
hydrophobicity due to water disorder from the displacements of the
amino acids.\\

\begin{figure}[H]
\begin{center}
\includegraphics[width=0.35\textwidth,angle=-90]{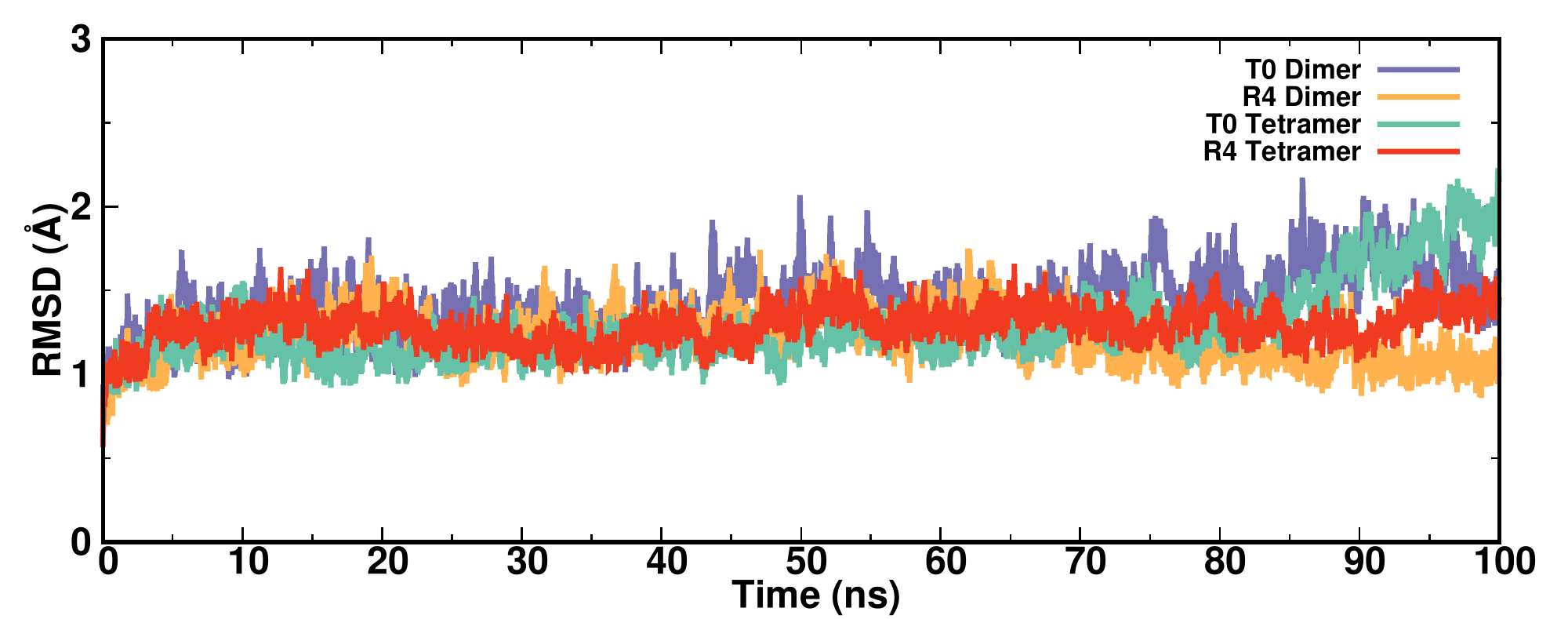}
\caption{The structural RMSD for the C$_{\alpha}$ atoms of the
  flexible T$_0$ and R$_4$ hemoglobin dimers and tetramers from 100 ns
  production runs. The RMSD is smaller for the T$_0$ tetramer than for
  the T$_0$ dimer, but for R$_4$, the tetramer RMSD is larger than for
  the dimer. However, it is noted that X-ray structures to compare
  with are available only for T$_0$ and R$_4$ but not for S1 of either
  of the tetramers.}
\label{fig:rmsd}
\end{center}
\end{figure}

\noindent
Figure \ref{fig:rmsd} reports the root mean squared deviation (RMSD)
for the C$_{\alpha}$ atoms for flexible tetrameric T$_0$ (cyan) and
R$_4$ (red) together with that for S1 of T$_0$ (blue) and R$_4$
(orange). For the most part all RMSD values are well below 2 \AA\/
except for occasional, short stochastic fluctuations for S1 of
T$_0$. Overall, the fluctuations for the tetrameric systems are
smaller than those for the dimers, except between 90 and 100 ns, where
the T$_0$ tetramer results are larger than those for the dimer. The
increase in RMSD for the T$_0$ tetramer after $\sim 90$ ns is
reminiscent of earlier findings that depending on water box size the
T$_0$ structure can become unstable.\cite{MM.hb:2018} For the isolated
$(\alpha \beta)$ subunits (S1 or S2 in the present case) of human
tetrameric Hb it should be noted that there is no experimental
information on their thermodynamic stability. Simulations for a
separate subunit (S1 or S2) were carried out primarily to be able to
quantify changes between the local hydrophobicity and water exposure
for the subunits vs. the tetramer at the relevant association
interface. For rigid tetrameric and dimeric Hb these simulations are
well-defined whereas for the flexible $(\alpha \beta)$ subunits the
results cannot be independently validated vis-a-vis experiments and
need to be considered with caution. The T$_0$ and R$_4$ tetramer
structures were found to be stable in the $90^3$ \AA\/$^3$ box for
about 500 ns for T$_0$ and no decay was reported for the R$_4$
state.\cite{MM.hb:2018}\\

\subsection{Local Hydrophobicity from Simulations with Rigid and Flexible Proteins}

\begin{figure}[h]
  \centering
  \includegraphics[scale=0.6]{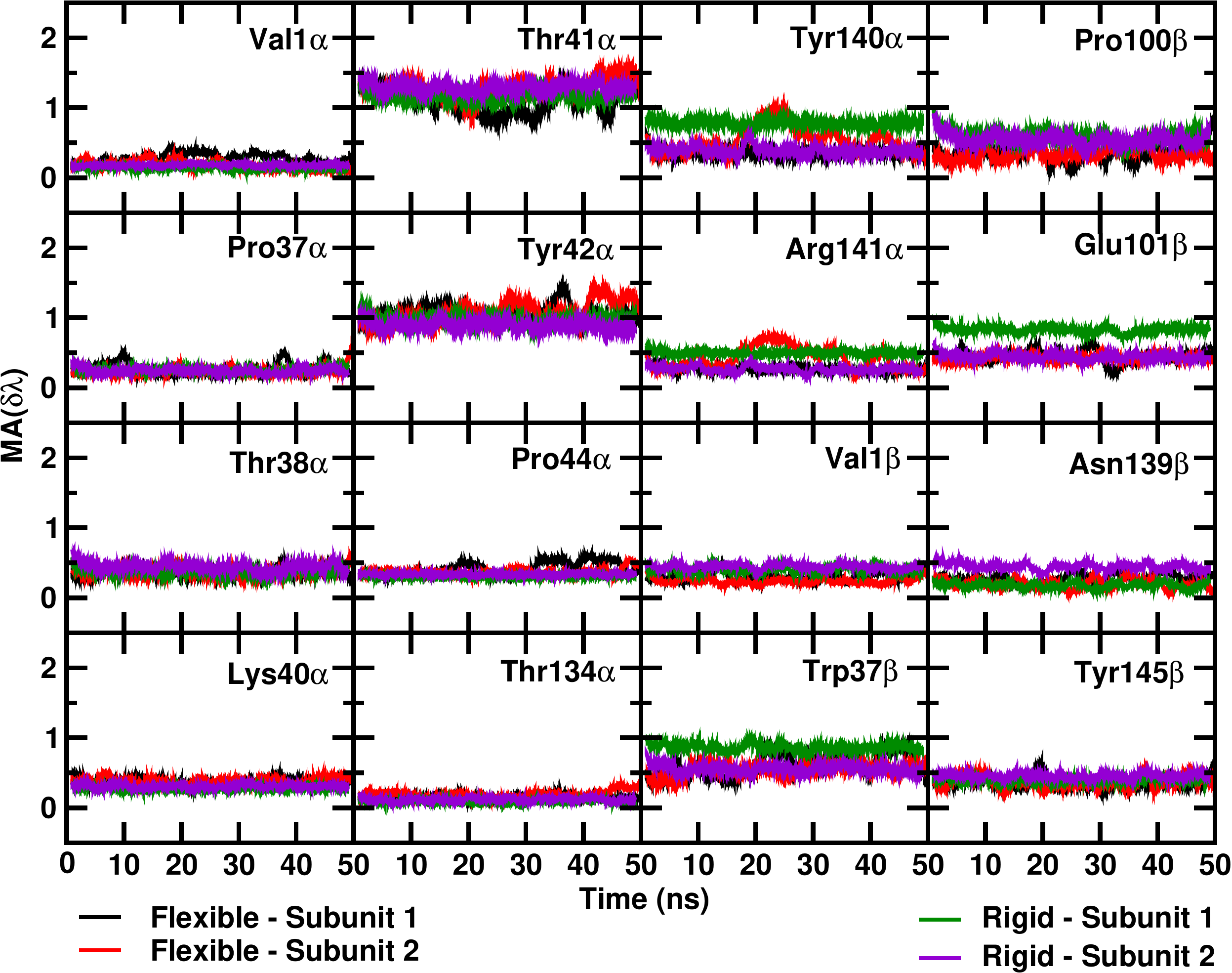}
\caption{LH for flexible and rigid T$_0$ Hb tetramer from 50 ns
  simulations: The two different subunits are represented in black
  ($\alpha_1\beta_1$) and red ($\alpha_2\beta_2$) for the flexible and
  in green ($\alpha_1\beta_1$) and violet ($\alpha_2\beta_2$) for the
  rigid tetramer. A window of 200 points was used for the running
  average.}
\label{fig:lh.t0}
\end{figure}

\noindent
Results for LH$(t)$ of the residues studied for rigid and flexible
tetrameric T$_0$ are reported in Figure \ref{fig:lh.t0} for S1
($\alpha_1 \beta_1$) in (green and black) and S2 ($\alpha_2 \beta_2$)
in (violet and red) from simulations 50 ns in length. The LH$(t)$ for
rigid and flexible tetrameric R$_4$ are shown in Figure
\ref{sifig:lh.r4}. For the rigid tetramers the time series for many of
the S1 and S2 residues are nearly identical. This is particularly true
for R$_4$ for which the only slight difference occurs for residue
Tyr140$\alpha$. For T$_0$ more differences arise, including
Thr41$\alpha$, Tyr42$\alpha$, Tyr140$\alpha$, Arg141$\alpha$,
Trp37$\beta$, Pro100$\beta$, Glu101$\beta$, and Asn139$\beta$ which
reflect differences in the experimentally published structure of the
two subunits S1 and S2 for the two tetramers. The RMSD between S1 and
S2 for the C$_\alpha$ atoms is 0.32 \AA\/ for T$_0$ (2DN2) and 0.001
\AA\/ for R$_4$ (2DN3). For 2DN2 the differences between S1 and S2
arise from the random coils connecting the alpha helices. Hence, for a
symmetric tetramer with superimposing subunits S1 and S2 (R$_4$)
LH$(t)$ and $P({\rm LH})$ are nearly identical for symmetry-equivalent
residues whereas for slight structural differences between S1 and S2
(T$_0$) some differences are observed.\\

\noindent
For the flexible T$_0$ tetramer (black and red traces for S1 and S2 in
Figure \ref{fig:lh.t0}) it is noted that almost all residues have
near-identical average values for LH. This is even the case even for
residues for which LH$(t)$ differed for the rigid tetramer. Examples
include residues Tyr140$\alpha$, Arg141$\alpha$, Trp37$\beta$,
Pro100$\beta$, and Glu101$\beta$. For these residues the dynamics
essentially ``symmetrizes'' the two dimers. Two classes of residues
can be distinguished: those for which the average $<{\rm LH}>$ for the
rigid and the flexible tetramer is nearly the same and others for
which the average differs due to the dynamics. Residues for which the
average hydrophobicity for rigid and flexible tetramer is equal,
include Val1$\alpha$, Pro37$\alpha$, Thr38$\alpha$, (Lys40$\alpha$),
Thr134$\alpha$, and Tyr145$\beta$. For Thr41$\alpha$, Tyr42$\alpha$,
Glu101$\beta$, and Asn139$\beta$ the differences between rigid and
flexible tetramers are particularly large. They can reach values of up
to 0.5 units for LH. Typically, including dynamics leads to a shift
towards lower values of LH (less hydrophilic); examples are
Tyr140$\alpha$, Arg141$\alpha$, Trp37$\beta$, and Glu101$\beta$.\\

\begin{figure}[h]
  \centering
  \includegraphics[scale=0.5]{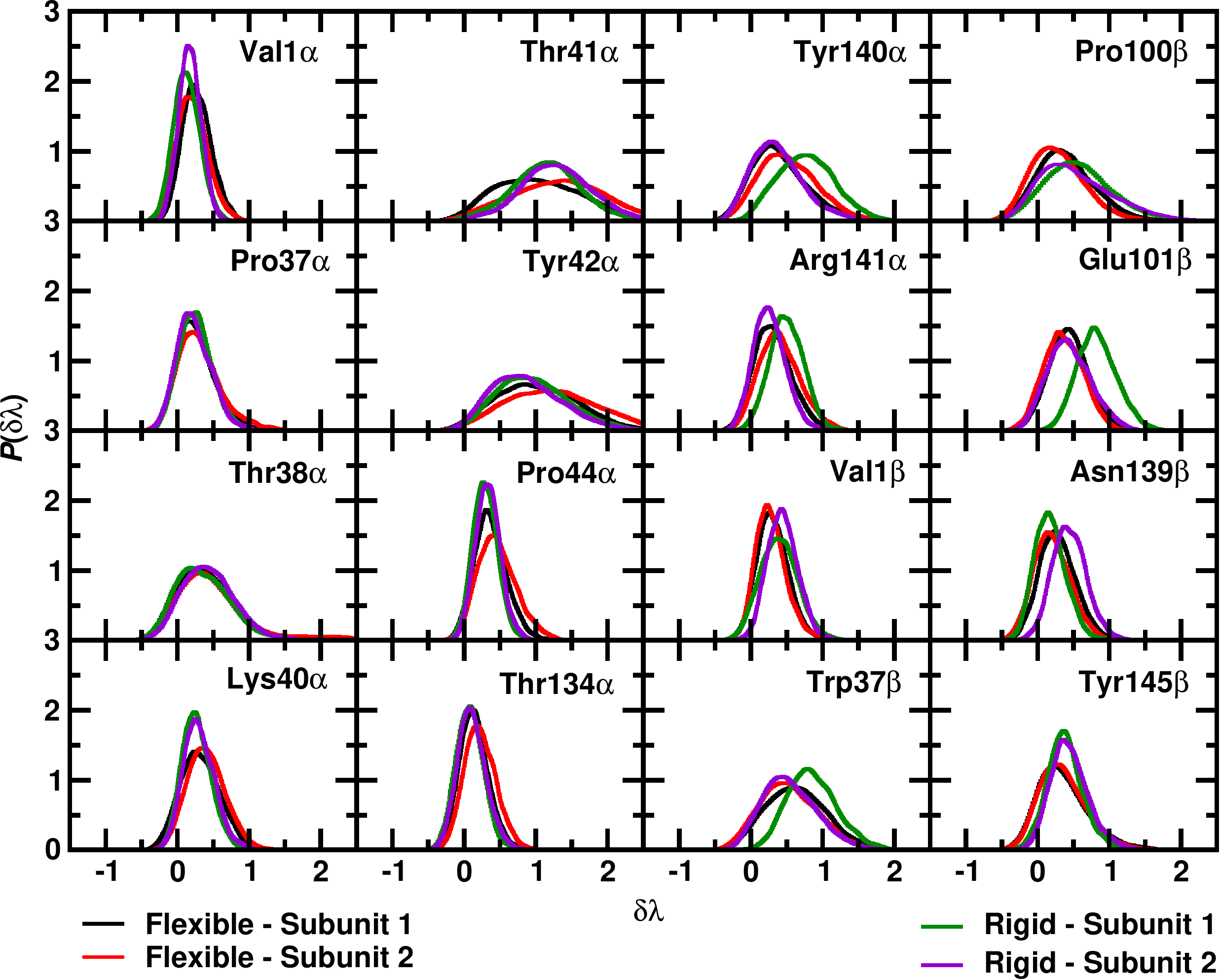}
\caption{$P({\rm LH})$ for flexible and rigid T$_0$ tetramer from 50
  ns simulations. The two different subunits are represented in black
  ($\alpha_1\beta_1$) and red ($\alpha_2\beta_2$) for flexible and in
  green ($\alpha_1\beta_1$) and violet ($\alpha_2\beta_2$) for rigid
  tetramer.}
\label{fig:plh.t0}
\end{figure}

\noindent
The differences between rigid and flexible tetramers in the R$_4$
state are in the opposite direction from T$_0$. The dynamics shifts
the LHs to more positive values (more hydrophilic) for Thr38$\alpha$,
Tyr42$\alpha$, Arg141$\alpha$, Asn139$\beta$; see Figure
\ref{sifig:lh.r4}. For residues Thr38$\alpha$ and Tyr42$\alpha$ the
local hydrophobicities for rigid and flexible tetramer differ
most. Interestingly, in the case of flexible R$_4$ a few residues in
S1 and S2 behave slightly differently from each other. They include
Thr38$\alpha$, Tyr42$\alpha$, Tyr140$\alpha$, and Arg141$\alpha$. \\

\noindent
Figure \ref{fig:plh.t0} shows the probability distribution functions,
$P({\rm LH})$, of the local hydrophobicities for T$_0$ determined from
the time series in Figure \ref{fig:lh.t0}. The distributions $P$(LH)
display near-Gaussian (e.g. Pro37$\alpha$) to non-Gaussian
(e.g. Tyr42$\alpha$) shapes. For this reason it was decided to
consider the position of the maximum, max$P$(LH), instead of the
arithmetic mean in the following. Similar to what was found for
LH$(t)$ in Figure \ref{fig:lh.t0}, the distributions overlap for the
majority of residues. For residues Thr41$\alpha$, Tyr42$\alpha$,
Pro44$\alpha$, Tyr140$\alpha$, Trp37$\beta$, and Pro100$\beta$ there
are significant differences for the flexible tetramer and for
Tyr140$\alpha$, Arg141$\alpha$, Val1$\beta$, Trp37$\beta$,
Glu101$\beta$, and Asn139$\beta$ they differ significantly for the
rigid tetramer. The probability distributions for R$_4$, reported in
Figure \ref{sifig:plh.r4}, are overlapping for all residues if the
protein structure is frozen, except for Tyr140$\alpha$ and
Arg141$\alpha$, for which very slight differences are found. In
contrast to that and to the flexible T$_0$ tetramer, dynamics leads to
some differences between symmetry-equivalent residues in R$_4$; they
include residues Tyr42$\alpha$ (pronounced), Tyr140$\alpha$,
Arg141$\alpha$, Trp37$\beta$, and Glu101$\beta$.\\

\noindent
The total LH for rigid T$_0$ and R$_4$ tetramers changes from
LH(T$_0$)$=13.7$ to LH(R$_4$)$=10.0$, i.e. from more hydrophilic to
less hydrophilic, see Table \ref{tab:tab1}. This is reversed if the
two tetramers are flexible for which LH(T$_0$)$=12.5$ and
LH(R$_4$)$=13.3$ as shown in Table \ref{tab:tab2}. Hence, flexibility
of the protein structure influences the magnitude of LH because
structural changes allow water exchange between bulk and the
protein-protein interfaces. \\

\noindent
Next, the total LH for frozen dimer and tetramer simulations is
considered. Formation of the tetramer causes some interfacial residues
to become buried compared with the dimers S1 and S2. This changes
their local hydrophobicity due to alterations in solvent access.  The
values for max$P$(LH) in Table \ref{tab:tab1} show that for rigid
T$_0$ the total change between the dimer and the tetramer for all the
residues analyzed here is $-6.78$ for S1 and $-7.90$ for S2 whereas
for R$_4$ it is $-9.34$ and $-8.88$, respectively. A positive number
in the change is associated with increased hydrophilicity. Hence, upon
S1/S2 association both tetramers' hydrophilicity decreases and the
total effect is larger for R$_4$ ($-18.22$) than for T$_0$
($-14.68$). If only residues at the $\alpha_1 \beta_2$ and $\alpha_2
\beta_1$ interfaces are considered (see Table \ref{tab:tabres}) the
total change between dimer and tetramer is $-1.33$ for T$_0$ (if
residue Asn139 is excluded because it points towards the channel, the
actual change is positive.)  and --13.5 for R$_4$. Hence, for the
T$_0$ tetramer the contribution of the interface is near-neutral
(LH$\sim 0$) but it is significantly hydrophobic (LH $= -13.5$) for
R$_4$.\\

\noindent
Next, the difference in LH between the tetramer and the dimer for
rigid T$_0$ and R$_4$ for each of the 16 residues is
considered. Figure \ref{sifig:corr_diff_rig} shows that all residues
appear approximately as pairs, as expected.  Upon association,
residues Thr41$\alpha$, Tyr42$\alpha$ and Trp37$\beta$ become more
hydrophilic for both, T$_0$ and R$_4$ ($+,+$ quadrant in Figure
\ref{sifig:corr_diff_rig}); residues Pro44$\alpha$ and Glu101$\beta$
become more hydrophilic for T$_0$ but less hydrophilic for R$_4$
($+,-$); Val1$\alpha$, Pro37$\alpha$, Thr38$\alpha$, Lys40$\alpha$,
Thr134$\alpha$, Val1$\beta$, Pro100$\beta$, Asn139$\beta$ and
Tyr145$\beta$ become less hydrophilic for both T$_0$ and R$_4$
($-,-$); and Tyr140$\alpha$ and Arg141$\alpha$ become more hydrophilic
for R$_4$ and less hydrophilic for T$_0$ ($-,+$). These results change
appreciably for the flexible dimer and tetramer simulations, see
Figure \ref{sifig:corr_diff_flex}. In this case, all differences are
in the upper right quadrant ($+,+$), which indicates that upon
association all residues become more hydrophilic for flexible T$_0$
and R$_4$. Figures \ref{sifig:rmsd-t0} and \ref{sifig:rmsd-r4} compare
the rigid with dynamically averaged structures for T$_0$ and R$_4$,
respectively; the RMSDs for the averaged structures are 1.65 \AA\/ for
T$_0$.  and 1.77 \AA\/ for R$_4$.  The differences are small, but for
both cases, it appears there is a small collapse in the averaged
structures.\\

\noindent
The local hydrophobicity for the individual residues in the dimer and
tetramer are compared for rigid and flexible proteins for T$_0$ and
R$_4$ in Figures \ref{sifig:corr_t0_rigid} to
\ref{sifig:corr_r4_flex}. As association of S1 and S2 to form the
tetramer leads to burying water-exposed parts of the protein for which
(${\rm LH} > 0$), the net effect of association is expected to be
reduced LH-values for residues involved in the association interface
in the tetramer; these are the residues in Table
\ref{tab:tabres}. This is largely what is observed for both T$_0$ and
R$_4$ in the rigid systems (Figures \ref{sifig:corr_t0_rigid} and
\ref{sifig:corr_r4_rigid}). There are a few exceptions for which the
LH-value in the tetramer is more positive than for the same residue in
S1 and S2. For T$_0$ (Figure \ref{sifig:corr_t0_rigid}) they are
residues Thr41$\alpha1$, Tyr42$\alpha1$, Trp37$\beta1$, Glu101$\beta1$
in S1 and S2 whereas for R$_4$ (Figure \ref{sifig:corr_r4_rigid}),
this is true for Tyr140$\alpha_1$ and Tyr140$\alpha_2$ and several
other residues in the lower left-hand corner (e.g. Trp37$\beta_2$),
though there is no relation to the results in Table
\ref{tab:tabres}. For the flexible systems (Figures
\ref{sifig:corr_t0_flex} and \ref{sifig:corr_r4_flex}), the LH values
are larger for the tetramer than the dimer for all residues in R$_4$
and for T$_0$ with the exception of Thr134$\alpha_1$, Val$\alpha_2$,
Val$\beta_2$ and Asn139$\beta_2$. The comparisons involving the
flexible dimer need to be considered with caution because there is no
experimental information on the structure and dynamics for isolated S1
in solution.\\

\noindent
Next, the effect of protein dynamics on the max$P$(LH) in the
tetramers is considered by comparing LH from rigid and flexible
simulations, see Figures \ref{sifig:corr_rigid_flex_t0} and
\ref{sifig:corr_rigid_flex_r4}. Overall, the position of the maxima
for all residues are approximately correlated in T$_0$ and highly
correlated in R$_4$.  There are some exceptions, namely
Thr41$\alpha(1,2)$, Tyr140$\alpha(1,2)$, Trp37$\beta(1,2)$, and
Glu101$\beta(1,2)$ for T$_0$ and Tyr42$\alpha(1,2)$, and to a lesser
extent Thr38$\alpha(1,2)$ and Tyr140$\alpha(1,2)$ for R$_4$. Residues
below the diagonal are more hydrophilic in the rigid than the flexible
protein whereas for those above the diagonal, protein dynamics
decreases their hydrophilicity. For T$_0$ (Figure
\ref{sifig:corr_rigid_flex_t0}) the exceptions are typically less
hydrophilic when dynamics is included whereas for R$_4$ (Figure
\ref{sifig:corr_rigid_flex_r4}) the opposite is the case.\\

\begin{table}[]
    \caption{Comparison of LH for T$_0$ and R$_4$ states of rigid
      dimers and tetramers. LH values are reported as the maximum of
      the distribution (max$P({\rm LH})$). Values in parentheses are
      the standard deviation. $\Delta$LH$_{\rm T_0}$ and
      $\Delta$LH$_{\rm R_4}$ refers to (tetramer - dimer)
      hydrophobicity. The labels Sum, Total and Global refer to the
      aggregate for the $\alpha$ and $\beta$ subunits, for S1 and S2,
      and for the global sum involving all LHs.}
    \label{tab:tab1}
    \centering
\begin{tabular}{l||rr|r||rr|r||r}
\toprule
Residue     & LH$_{\rm T_0}^{\rm tetra}$ & LH$_{\rm T_0}^{\rm dimer}$ & $\Delta$LH$_{\rm T_0}$ &
LH$_{\rm R_4}^{\rm tetra}$ & LH$_{\rm R_4}^{\rm dimer}$ & $\Delta$LH$_{\rm R_4}$ & $\Delta$LH$^{\rm tetra}_{\rm R_4 - T_0}$ \\
\hline
Val1$\alpha$1   & 0.12 (0.76) & 1.40 (0.44) & --1.28                   & 0.28 (0.68) & 0.42 (0.13) & --0.14                   & 0.16  \\
Pro37$\alpha$1  & 0.26 (0.58) & 1.01 (0.29) & --0.75                   & 0.29 (0.69) & 1.18 (0.21) & --0.89                   & 0.03  \\
Thr38$\alpha$1  & 0.21 (0.36) & 0.86 (0.31) & --0.65                   & 0.63 (0.32) & 0.76 (0.18) & --0.13                   & 0.42  \\
Lys40$\alpha$1  & 0.25 (0.68) & 0.34 (0.27) & --0.09                   & 0.22 (0.55) & 1.54 (0.20) & --1.32                   & --0.03 \\
Thr41$\alpha$1  & 1.20 (0.29) & 0.27 (0.23) & 0.93                    & 0.29 (0.47) & 0.23 (0.18) & 0.06                    & --0.91 \\
Tyr42$\alpha$1  & 0.83 (0.28) & 0.12 (0.18) & 0.71                    & 0.51 (0.36) & 0.17 (0.17) & 0.34                    & --0.32 \\
Pro44$\alpha$1  & 0.27 (0.80) & 0.13 (0.18) & 0.14                    & 0.24 (0.86) & 0.59 (0.12) & --0.35                   & --0.03 \\
Thr134$\alpha$1 & 0.09 (0.64) & 1.40 (0.37) & --1.31                   & 0.11 (0.54) & 0.27 (0.12) & --0.16                   & 0.02  \\
Tyr140$\alpha$1 & 0.79 (0.35) & 1.61 (0.24) & --0.82                   & 0.48 (0.35) & 0.29 (0.16) & 0.19                    & --0.31 \\
Arg141$\alpha$1 & 0.44 (0.58) & 1.52 (0.53) & --1.08                   & 0.28 (0.93) & 0.24 (0.18) & 0.04                    & --0.16 \\
\textbf{Sum:}   & 4.46        &  8.66       & --4.20                    &  3.33       & 5.69        & --2.36                   &  --1.13\\  
\midrule 
Val1$\beta$1    & 0.37 (0.52) & 1.92 (0.22) & --1.55                   & 0.38 (0.50) & 1.40 (0.25) & --1.02                   & 0.01  \\
Trp37$\beta$1   & 0.79 (0.40) & 0.00 (0.73) & 0.79                    & 0.31 (0.30) & 0.19 (0.14) & 0.12                    & --0.48 \\
Pro100$\beta$1  & 0.47 (0.26) & 0.73 (0.30) & --0.26                   & 0.02 (0.57) & 1.86 (0.19) & --1.84                   & --0.45 \\
Glu101$\beta$1  & 0.79 (0.50) & 0.36 (0.51) & 0.43                    & 0.14 (0.72) & 1.65 (0.18) & --1.51                   & --0.65 \\
Asn139$\beta$1  & 0.15 (0.63) & 1.81 (0.30) & --1.66                   & 0.36 (0.47) & 1.65 (0.26) & --1.29                   & 0.21  \\
Tyr145$\beta$1  & 0.37 (0.56) & 0.70 (0.36) & --0.33                   & 0.24 (0.57) & 1.68 (0.25) & --1.44                   & --0.13 \\
\textbf{Sum:}   & 2.94        &  5.52       & --2.58                   & 1.45       & 8.43         & --6.98                   & --1.49 \\  
\bottomrule                                                                                         
\textbf{Total S1:}   &7.40         &  14.18      & --6.78                   &  4.78      & 14.12        &  --9.34                  & --2.62 \\   
\midrule                                                
Val1$\alpha$2   & 0.15 (0.87) & 1.40 (0.44) & --1.25                   & 0.32 (0.65) & 0.42 (0.13) & --0.10                   & 0.17   \\
Pro37$\alpha$2  & 0.16 (0.57) & 1.01 (0.29) & --0.85                   & 0.30 (0.66) & 1.18 (0.21) & --0.88                   & 0.14   \\
Thr38$\alpha$2  & 0.37 (0.37) & 0.86 (0.31) & --0.49                   & 0.73 (0.32) & 0.76 (0.18) & --0.03                   & 0.36   \\
Lys40$\alpha$2  & 0.26 (0.64) & 0.34 (0.27) & --0.08                   & 0.21 (0.54) & 1.54 (0.20) & --1.33                   & --0.05  \\
Thr41$\alpha$2  & 1.25 (0.28) & 0.27 (0.23) & 0.98                    & 0.35 (0.47) & 0.23 (0.18) & 0.12                    & --0.90  \\
Tyr42$\alpha$2  & 0.80 (0.28) & 0.12 (0.18) & 0.68                    & 0.51 (0.39) & 0.17 (0.17) & 0.34                    & --0.29  \\
Pro44$\alpha$2  & 0.32 (0.79) & 0.13 (0.18) & 0.19                    & 0.27 (0.84) & 0.59 (0.12) & --0.32                   & --0.05  \\
Thr134$\alpha$2 & 0.09 (0.63) & 1.40 (0.37) & --1.31                   & 0.12 (0.57) & 0.27 (0.12) & --0.15                   & 0.03   \\
Tyr140$\alpha$2 & 0.30 (0.38) & 1.61 (0.24) & --1.31                   & 0.69 (0.33) & 0.29 (0.16) & 0.40                    & 0.39   \\
Arg141$\alpha$2 & 0.24 (0.62) & 1.52 (0.53) & --1.28                   & 0.40 (0.90) & 0.24 (0.18) & 0.16                    & 0.16   \\
\textbf{Sum:}   & 3.94        & 8.66        & --4.72                   & 3.90        & 5.69        & --1.79                   &--0.04  \\  
\midrule 
Val1$\beta$2    & 0.43 (0.65) & 1.92 (0.22) & --1.49                   & 0.35 (0.49) & 1.40 (0.25) & --1.05                   & --0.08  \\
Trp37$\beta$2   & 0.44 (0.36) & 0.00 (0.73) & 0.44                    & 0.27 (0.28) & 0.19 (0.14) & 0.08                    & --0.17  \\
Pro100$\beta$2  & 0.31 (0.28) & 0.73 (0.30) & --0.42                   & 0.08 (0.63) & 1.86 (0.19) & --1.78                   & --0.23  \\
Glu101$\beta$2  & 0.41 (0.45) & 0.36 (0.51) & 0.05                    & 0.14 (0.70) & 1.65 (0.18) & --1.51                   & --0.27  \\
Asn139$\beta$2  & 0.39 (0.57) & 1.81 (0.30) & --1.42                   & 0.27 (0.47) & 1.65 (0.26) & --1.38                   & --0.12  \\
Tyr145$\beta$2  & 0.36 (0.53) & 0.70 (0.36) & --0.34                   & 0.23 (0.62) & 1.68 (0.25) & --1.45                   & --0.13  \\              
\textbf{Sum:}   & 2.34        & 5.52        & --3.18                   & 1.34        & 8.43        & --7.09                   & --1.00  \\   
\midrule
\textbf{Total S2:}     & 6.28        & 14.18       & --7.90                   & 5.24        & 14.12       &--8.88                   & --1.04   \\
\bottomrule                                                                                        
\textbf{Global S1$+$S2:}   & 13.68        & 28.36       & --14.68                   & 10.02        & 28.24       &--18.22                   & --3.66   \\   
\bottomrule
\end{tabular}
\end{table}

\begin{table}[]
\caption{Comparison of LH for T$_0$ and R$_4$ states of flexible
  dimers and tetramers. LH values are reported as the maximum of the
  distribution (max$P({\rm LH})$). Values in parentheses are the
  standard deviation. $\Delta$LH$_{\rm T_0}$ and $\Delta$LH$_{\rm
    R_4}$ refers to (tetramer - dimer) hydrophobicity. The labels Sum,
  Total and Global refer to the aggregate for the $\alpha$ and $\beta$
  subunits, for S1 and S2, and for the global sum involving all LHs.}
    \label{tab:tab2}
    \centering
\begin{tabular}{l||rr|r||rr|r||r}
\toprule
Residue     & LH$_{\rm T_0}^{\rm tetra}$ & LH$_{\rm T_0}^{\rm dimer}$ & $\Delta$LH$_{\rm T_0}$ &
LH$_{\rm R_4}^{\rm tetra}$ & LH$_{\rm R_4}^{\rm dimer}$ & $\Delta$LH$_{\rm R_4}$ & $\Delta$LH$^{\rm tetra}_{\rm R_4 - T_0}$ \\
\hline
Val1$\alpha$1   & 0.21(0.67) & 0.18(0.62)  & 0.03  & 0.41(0.57) & 0.21(0.66)  & 0.20 & 0.20  \\
Pro37$\alpha$1  & 0.17(0.5)  & 0.12(0.68)  & 0.05  & 0.34(0.61) & 0.09(0.71)  & 0.25 & 0.17  \\
Thr38$\alpha$1  & 0.34(0.35) & 0.18(0.56)  & 0.16  & 0.98(0.31) & 0.18(0.59)  & 0.80 & 0.64  \\
Lys40$\alpha$1  & 0.25(0.49) & 0.05(0.63)  & 0.20  & 0.27(0.55) & 0.03(0.64)  & 0.24 & 0.02  \\
Thr41$\alpha$1  & 0.95(0.23) & 0.13(0.67)  & 0.82  & 0.52(0.43) & 0.12(0.71)  & 0.40 & --0.43 \\
Tyr42$\alpha$1  & 0.86(0.24) & 0.39(0.5)   & 0.47  & 1.25(0.25) & 0.18(0.58)  & 1.07 & 0.39  \\
Pro44$\alpha$1  & 0.32(0.61) & 0.28(0.62)  & 0.04  & 0.29(0.78) & 0.24(0.70)  & 0.05 & --0.03 \\
Thr134$\alpha$1 & 0.13(0.69) & 0.16(0.6)   & --0.03 & 0.2 (0.55) & 0.07(0.65)  & 0.13 & 0.07  \\
Tyr140$\alpha$1 & 0.28(0.36) & 0.22(0.56)  & 0.06  & 0.5 (0.38) & 0.08(0.59)  & 0.42 & 0.22  \\
Arg141$\alpha$1 & 0.28(0.51) & 0.19(0.69)  & 0.09  & 0.52(0.53) & 0.17(0.68)  & 0.35 & 0.24  \\
\textbf{Sum:}      & 3.79       & 1.90        & 1.89  & 5.28       & 1.37        & 3.91 & 1.49  \\
\midrule 
Val1$\beta$1   & 0.27(0.61) & 0.29(0.60)  & --0.02 & 0.36(0.48) & 0.19(0.61)  & 0.17 & 0.09  \\
Trp37$\beta$1  & 0.63(0.32) & 0.13(0.71)  & 0.50  & 0.25(0.37) & 0.09(0.75)  & 0.16 & --0.38 \\
Pro100$\beta$1 & 0.30(0.35) & --0.14(0.62) & 0.44  & 0.10(0.57) & --0.02(0.61) & 0.12 & --0.20 \\
Glu101$\beta$1 & 0.44(0.50) & --0.07(0.62) & 0.51  & 0.15(0.58) & --0.04(0.57) & 0.19 & --0.29 \\
Asn139$\beta$1 & 0.25(0.54) & 0.17(0.54)  & 0.08  & 0.43(0.38) & 0.22(0.52)  & 0.21 & 0.18  \\
Tyr145$\beta$1 & 0.25(0.38) & --0.01(0.63) & 0.26  & 0.28(0.56) & 0.07(0.79)  & 0.21 & 0.03  \\
\textbf{Sum:}      & 2.14       & 0.37        & 1.77  & 1.57       & 0.51        & 1.06 & --0.57 \\
\midrule
\textbf{Total S1:}      & 5.93       & 2.27        & 3.66  & 6.85       & 1.88        & 4.97 & 0.92  \\
\bottomrule
Val1$\alpha$2   & 0.14(0.64) & 0.18(0.62)  & --0.04 & 0.37(0.57) & 0.21(0.66)  & 0.16 & 0.23  \\
Pro37$\alpha$2  & 0.23(0.47) & 0.12(0.68)  & 0.11  & 0.36(0.59) & 0.09(0.71)  & 0.27 & 0.13  \\
Thr38$\alpha$2  & 0.33(0.29) & 0.18(0.56)  & 0.15  & 0.85(0.29) & 0.18(0.59)  & 0.67 & 0.52  \\
Lys40$\alpha$2  & 0.37(0.52) & 0.05(0.63)  & 0.32  & 0.25(0.46) & 0.03(0.64)  & 0.22 & --0.12 \\
Thr41$\alpha$2  & 1.41(0.21) & 0.13(0.67)  & 1.28  & 0.45(0.41) & 0.12(0.71)  & 0.33 & --0.96 \\
Tyr42$\alpha$2  & 1.16(0.21) & 0.39(0.50)  & 0.77  & 0.75(0.30) & 0.18(0.58)  & 0.57 & --0.41 \\
Pro44$\alpha$2  & 0.42(0.52) & 0.28(0.62)  & 0.14  & 0.28(0.63) & 0.24(0.70)  & 0.04 & --0.14 \\
Thr134$\alpha$2 & 0.18(0.61) & 0.16(0.60)  & 0.02  & 0.15(0.59) & 0.07(0.65)  & 0.08 & --0.03 \\
Tyr140$\alpha$2 & 0.36(0.34) & 0.22(0.56)  & 0.14  & 0.49(0.38) & 0.08(0.59)  & 0.41 & 0.13  \\
Arg141$\alpha$2 & 0.37(0.48) & 0.19(0.69)  & 0.18  & 0.54(0.62) & 0.17(0.68)  & 0.37 & 0.17  \\
\textbf{Sum:}      & 4.97       & 1.90        & 3.07  & 4.49       & 1.37        & 3.12 & --0.48 \\
\midrule 
Val1$\beta$2   & 0.23(0.64) & 0.29(0.60)  & --0.06 & 0.39(0.51) & 0.19(0.61)  & 0.20 & 0.16  \\
Trp37$\beta$2  & 0.46(0.34) & 0.13(0.71)  & 0.33  & 0.47(0.34) & 0.09(0.75)  & 0.38 & 0.01  \\
Pro100$\beta$2 & 0.17(0.36) & --0.14(0.62) & 0.31  & 0.17(0.54) & --0.02(0.61) & 0.19 & 0.00  \\
Glu101$\beta$2 & 0.30(0.49) & --0.07(0.62) & 0.37  & 0.13(0.57) & --0.04(0.57) & 0.17 & --0.17 \\
Asn139$\beta$2 & 0.15(0.53) & 0.17(0.54)  & --0.02 & 0.47(0.38) & 0.22(0.52)  & 0.25 & 0.32  \\
Tyr145$\beta$2 & 0.30(0.42) & --0.01(0.63) & 0.31  & 0.28(0.57) & 0.07(0.79)  & 0.21 & --0.02 \\
\textbf{Sum:}      & 1.61       & 0.37        & 1.24  & 1.91       & 0.51        & 1.40 & 0.30  \\
\midrule
\textbf{Total S2:}      & 6.58       & 2.27     & 4.31  &   6.40    &    1.88     &4.52 & --0.18  \\
\bottomrule
\textbf{Global S1$+$S2:}      & 12.51       & 4.54        & 7.97  & 13.25       & 3.76        & 9.49 & 0.74\\
\bottomrule
\end{tabular}
\end{table}

\subsection{Local Hydrophobicity versus Solvent Accessible Surface Area}
The solvent exposure of buried and exposed interfacial residues for
tetrameric T$_0$ and R$_4$ was determined by computing the solvent
accessible surface area (SASA) for the two available high quality
X-ray structures.\cite{chothia:1985} Thus, the most direct and
meaningful comparison in the present context with these results is to
use LH from simulations in which the protein is rigid. The analysis
found that the deoxy state (T$_0$) buries 2620 \AA\/$^2$ of surface,
700 \AA\/$^2$ more than the oxy (R$_4$) state of Hb for the $\alpha_1
\beta_2$ and $\alpha_2 \beta_1$ interfaces.\\

\begin{figure}
  \centering
  \includegraphics[scale=1.3]{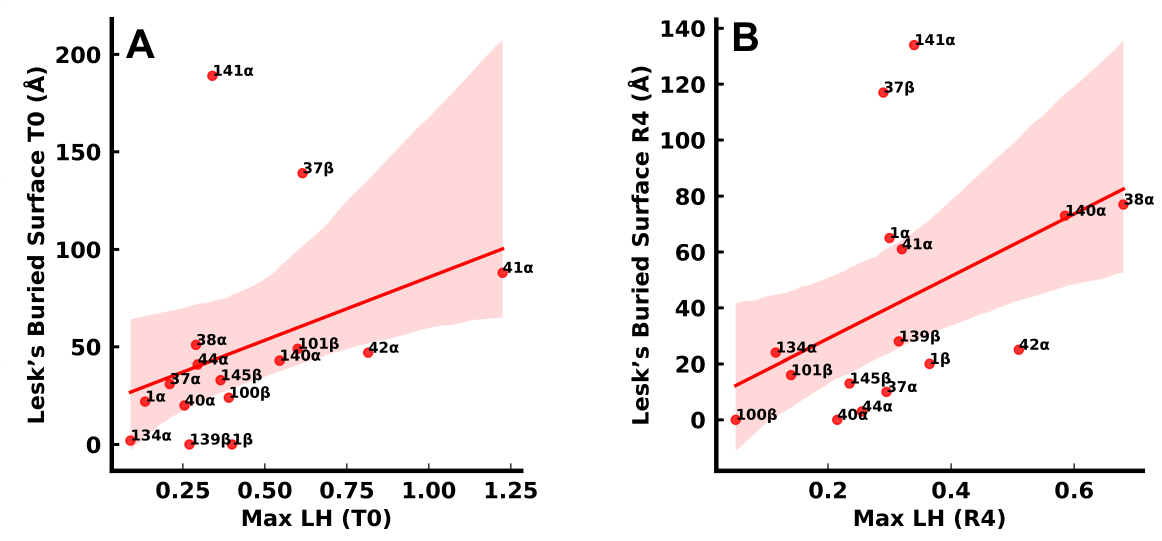}
\caption{Comparison between the averaged Max$P({\rm LH})$ for every
  residue and buried surface for the rigid tetramers. Panel A for
  T$_0$: The correlation coefficient is 0.36 and the shaded area
  indicates the 95 \% confidence interval. Panel B for R$_4$: The
  correlation coefficient is 0.44 and the shaded area indicates the 95
  \% confidence interval.}
\label{fig:lesk}
\end{figure}

\begin{figure}[h]
  \centering
  \includegraphics[scale=0.4]{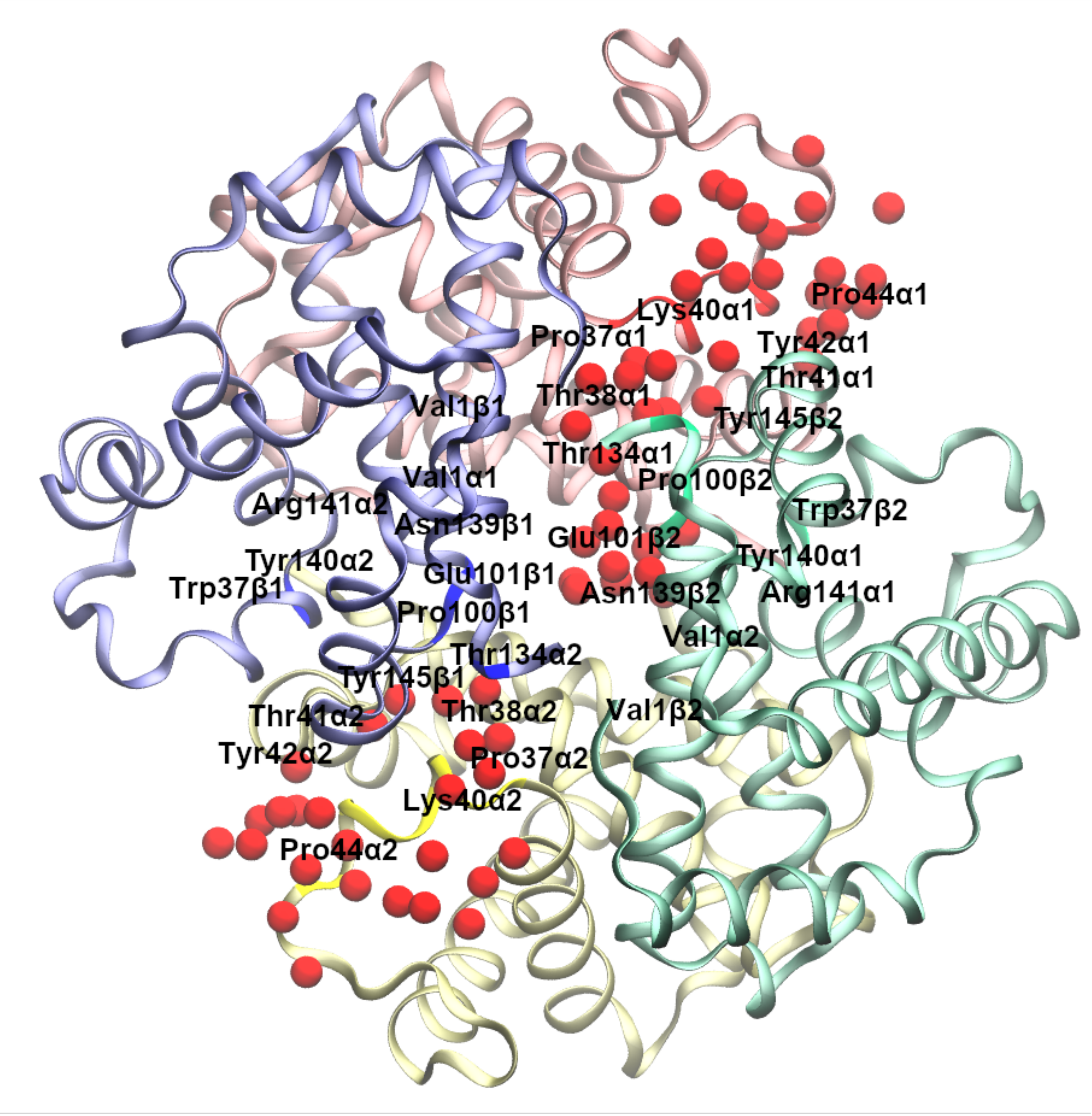}
  \caption{For R$_4$, the water molecules (red spheres) within 3 \AA\/
    of any residue identified by the ticks in Table \ref{tab:tabres}
    as being at the $\alpha_1 , \beta_2$/$\alpha_2 , \beta_1$
    interface with relevant residues labelled. The blue and green
    secondary structures refer to S1 and S2 and the relevant residues
    are labelled.}
  \label{fig:figinter}
\end{figure}

\noindent
This decrease in buried surface for the T$_0$$\rightarrow$R$_4$
transition correlates with the increased stability of the T- versus
the R-state based on the result of Chothia.\cite{chothia:1985} The
comparison between the local hydrophobicities from the rigid protein
tetramer simulations and the buried surface from the
literature\cite{chothia:1985} is reported in Figures \ref{fig:lesk}
and \ref{fig:lhavg.vs.lesk}. Figure \ref{fig:lesk} compares the buried
surface and max$P$(LH) for the T$_0$ (panel A) and R$_4$ states (panel
B) whereas Figure \ref{fig:lhavg.vs.lesk} provides a comprehensive
view of all available data. In general, increased max$P$(LH)
correlates with larger surface buried for both T$_0$ and R$_4$. For
both, the T- and the R-states there is a mild (for T$_0$, $R^2 =
0.36$) to a somewhat stronger (for R$_4$, $R^2 = 0.44$) correlation
between max$P$(LH) and the amount of buried surface.  Figure
\ref{fig:lesk} indicates that larger values of buried surface
correspond to larger values of LH. Since large values of buried
surface indicate that there is significant hydrophobic
stabilization\cite{chothia:1974} and positive values of LH indicate a
hydrophilic environment, Figure \ref{fig:lesk} and the results given
above point to a weak anticorrelation between SASA and LH. If large
values of buried surface in a protein is interpreted as ``the
probability to find water in these areas is low'' then the simulations
as per Figure \ref{fig:figinter} show that this is not the case: water
can access such areas even for rigid T$_0$ and R$_4$. Figures
\ref{fig:figinter} and \ref{sifig:t0water} show water molecules within
3 \AA\/ of any residue at the $\alpha_1 \beta_2$ and $\alpha_2
\beta_1$ association interfaces for rigid R$_4$ (77 waters) and T$_0$
(62 waters), respectively. These water molecules can be quite strongly
bound with lifetimes of several nanoseconds due to the rigidity of the
protein.\\

\noindent
There are two pronounced outliers for both analyses (buried surface
and LH), which are Arg141$\alpha$ and Trp37$\beta$; see Figure
\ref{fig:lesk}. The corresponding radial distribution functions are
reported in Figure \ref{sifig:gr}. For Trp37$\beta$ the average of the
maxima of LH, P(LH), for Trp37$\beta_1$ and Trp37$\beta_2$ is 0.62 for
T$_0$ and 0.30 for R$_4$ (Table 2), whereas the $g(r)$ are close to
one another (see red traces in Figure \ref{sifig:gr}). Hence, the
difference in the maxima of LH for Trp37$\beta_1$ and $\beta_2$ is
most likely due to water orientation around the two residues, although
the effect is small. This conclusion follows from the fact that the
$g(r)$, which probes only water presence, are similar and the max
$P$(LH), which probes both presence ($g(r)$) and orientation,
differ. For Arg141$\alpha$ the average of the maxima of $P({\rm LH}) =
0.34$ for both T$_0$ and R$_4$ whereas the $g(r)$ is much larger for
R$_4$ than for T$_0$ (see blue traces in Figure \ref{sifig:gr}). As
stated above, LH quantifies both the presence and the orientation of
solvent, while $g(r)$ only describes the presence of it. Consequently,
the findings that the max P(LH) are the same for Arg141 imply that for
R$_4$ the solvent molecules are unfavourably oriented for effective
protein-solvent interactions.\\

\noindent
The analysis based on LH for the rigid tetramers in their T$_0$ and
R$_4$ states shows that for the residues considered here, LH is larger
for T- than for R-Hb (13.7 vs. 10.0); i.e., LH is larger as a
(positive) number which means more hydrophilic. For S1 the total LH
for T$_0$ is 7.40 compared with 4.78 for R$_4$ and for S2 they are
6.28 and 5.24, respectively. If only residues at the $\alpha_1
\beta_2$ and $\alpha_2 \beta_1$ interfaces are considered (see Table
\ref{tab:tabres}), the total LH is 10.1 for T$_0$ as compared with 6.0
for R$_4$. Hence, T$_0$ appears to be more hydrophilic than R$_4$ when
measured by LH, again in disagreement with experiment.\\

\section{Discussion and Conclusion}
The present work uses local hydrophobicity as a physically based
measure for solvent exposure (and solvent structure) around hemoglobin
(Hb) to determine local hydrophilicity (LH$> 0.5$) and local
hydrophobicity (LH$< 0.5$). For rigid tetrameric Hb in its T$_0$ and
R$_4$ states it is found that the position of the maximum of the
distribution, $P$(LH), is mildly correlated with the more
conventionally used measure of ``buried surface'' and therefore mildly
anticorrelated with the solvent accessible surface area, see Figure
\ref{fig:lesk}. Large values of buried surface - i.e. hydrophobic
stabilization - for a given residue correlate with large values of LH,
the hydrophilicity; this is the inverse of the correlation that would
indicate that LH and buried surface measure corresponding
quantities.\\

\noindent
It was previously concluded\cite{chothia:1974} that larger areas of
buried surface correlate with increased hydrophobicity and
stabilization of the corresponding conformational
substate. Specifically, it was argued for Hb that the larger buried
surfaces for the $\alpha_1 \beta_2$ and $\alpha_2 \beta_1$ association
interfaces for T- versus R-states (2620 \AA\/$^3$ vs. 1920 \AA\/$^3$)
contribute significantly to the experimentally observed thermodynamic
stabilization of the T-state relative to the R-state. The total LH for
the residues at the relevant $\alpha_1 \beta_2$ and $\alpha_2 \beta_1$
interfaces indicate that these regions are more hydrophilic (max
$P$(LH) = 9.8) for the T-state compared with 5.9 for the R-state.
Several factors may contribute to this result. First, analysis of SASA
delineates hydrophobic regions where water access is expected to be
difficult and rare. However, even for rigid R$_4$ (see Figure
\ref{fig:figinter}), water is found to penetrate into such
regions. The existence of ordered water molecules at protein-protein
interfaces is quite general. For example, this has also been reported
for scapharca dimeric Hb.\cite{zhou:2003,MM.hb:2016} For this system,
the interface contains between 15 and 20 water molecules. Furthermore,
hydration of the Hb interface was found to change in the T/R
transition; experiments reported an increase by 60 water molecules in
the transition.\cite{colombo:1992}. Moreover, LH is sensitive not only
to the hydrophobic areas between residues forming the interface but
also to water access from the outside. In addition, LH depends on both
the presence and orientation of water molecules. Hence a single,
well-ordered water molecule may lead to values of LH that indicate a
hydrophilic nature (LH $\gtrsim 0.5$) of the residue if it is
optimally oriented ($\cos{\theta} \sim 1$, see Methods). Similarly,
multiple water molecules may result in a hydrophobic interface,
LH$\lesssim 0$, if they are unfavourably oriented.\\

\noindent
For the difference in LH between tetramer and dimer in both
conformational substates it is found that some residues are
surprisingly hydrophilic in the tetramer compared with the dimer, see
Figures \ref{sifig:corr_diff_rig} and \ref{sifig:corr_diff_flex}.
Thr41$\alpha$ is a typical example: for rigid tetramers and dimers the
average difference in Max$P({\rm LH})$ for T$_0$ is $\sim 1.0$ (1.18
for tetramer and 0.13 for dimer, see Table \ref{tab:tab1}) whereas for
R$_4$ it is $\sim 0.37$ (0.49 vs. 0.12). For the same residue, the
radial distribution functions for rigid R$_4$ in the dimer and the
tetramer overlap up to a separation of $\sim 3$ \AA\/ and differ
beyond, see green and blue traces in Figure \ref{sifig:gr3}. If the
proteins are flexible the Max$P({\rm LH})$ for the dimers in both
conformational substates only change marginally compared with the
rigid dimers and the values for Max$P({\rm LH})$ still indicate a
hydrophobic character (all Max$P({\rm LH}) < 0.5$). For flexible R$_4$
the average Max$P({\rm LH})$ increases somewhat (by 0.1) compared with
rigid R$_4$ and the corresponding radial distribution functions
(orange and red traces in Figure \ref{sifig:gr3}) show that for both
the dimer and the tetramer water can now penetrate closer to the
residue, for which $g(r)$ has a new local maximum for separations
smaller than 2 \AA\/ whereas the limiting value is reached at $\sim 5$
\AA\/. Since $g(r)$ for bulk water corresponds to a limit of one, the
smaller limiting values result from the presence of the protein. For
flexible T$_0$ the average Max$P({\rm LH})$ remains around 1.25 but
becomes more asymmetric with Max$P({\rm LH})$ larger for S2 than for
S1. The radial distribution function (Figure \ref{sifig:gr2}) for the
flexible tetramer only shows a faint density for separations of 2
\AA\/ (red trace) whereas for the flexible dimer water can access 2
\AA\/ more readily (orange trace). However, in this case, the limiting
value is reached for shorter separations ($\sim 4$ \AA\/) for the
tetramer compared with much larger values (greater than 7 \AA\/) for
the dimer.\\

\noindent
Finally, it is of interest to consider the present findings in the
general context of ``allostery''. The term - Greek for ``other site''
- used in the context of controlling cellular function at a molecular
level, was introduced in 1961 to describe ``interaction at a
distance'' involving two (or multiple) binding sites in a
protein.\cite{monod:1961} This model evolved into the celebrated
``Monod-Wyman-Changeux'' (MWC) model for allostery.\cite{monod:1965}
Historically, the concept was introduced even earlier, by Pauling, who
had proposed a model to explain positive cooperativity in binding of
molecular oxygen to hemoglobin.\cite{pauling:1935} This model was the
basis for an alternative view of cooperativity, now referred to as the
``Koshland-Nemethy-Filmer'' (KNF) model.\cite{koshland:1966} Applied
to Hb, the KNF model involves multiple conformational substates with
varying strengths for subunit interactions and the possibility for
intermediate states whereas the MWC is based on a concerted mechanism
for allostery at the quaternary level. The MWC model and its
elaborations are now accepted as the mechanism of cooperativity in
hemoglobin.\cite{szabo:1972}\\

\noindent
In conclusion, the present work introduces local hydrophobicity (LH)
as a meaningful and physically motivated measure for water exposure of
conformational substates in proteins. LH is anticorrelated with SASA
when both are measured for rigid Hb. Interestingly for flexible Hb, LH
correlates with the rigid SASA values. Overall, it appears that LH and
SASA measure different aspects of hydration. Since hydration is shown
to be important for protein function, it is essential for
allostery. \\

\section{Methods}
\subsection{Molecular Dynamics Simulation}
Molecular Dynamics (MD) simulations for rigid and flexible T$_0$ and
R$_4$ hemoglobin tetramer and for rigid and flexible subunits S1 were
performed using the CHARMM36\cite{charmm36:2010} force field with the
TIP3P water model\cite{jorgensen-tip3p} in a cubic box of size
$(90.0)^3$ \AA\/$^3$. The initial structures are the 2DN2 (T$_0$) and
2DN3 (R$_4$) structures\cite{tame:2006} solvated in a $90^3$ \AA\/$^3$
water box. All simulations were carried out for the protein frozen in
its X-ray conformation and for flexible Hb (``regular MD''). The local
hydrophobicity (LH) was analyzed for residues at the dimer-dimer
interface, whose buried surface area changes significantly between
T$_0$ and R$_4$ tetramer as reported by Lesk et al.\cite{chothia:1985}
(see their Table 1). The OpenMM implementation\cite{openMM} of C36 was
used together with CMAP corrections\cite{cmap06} for these
simulations. Electrostatic interactions were treated with the particle
mesh Ewald method\cite{Darden1993} with grid size spacing of 1 \AA\/,
characteristic reciprocal length $\kappa = 0.34$ \AA\/$^{-1}$, and
interpolation order 6. The simulations were run for 100 ns for both
T$_0$ and R$_4$ flexible and rigid dimers and for the flexible
tetramers and for 50 ns for the rigid tetramers. The LH-analysis
reports the maximum of the probability distribution $P({\rm LH})$
(i.e. its ``mode'') because several of the distributions were found to
be non-Gaussian. For the "rigid" simulations all protein atoms were
frozen at their positions in accord with the PDB structures. With 50
ns of dynamics for the two rigid tetramers the distributions, $P({\rm
  LH})$, were converged. This was verified by separately computing
$P({\rm LH})$ from the first and second 25 ns of the
simulation. Superimposing these distributions showed that they were
identical.\\

\subsection{Analysis of Aqueous Interfacial Structure}
The hydration structure around Hb was characterized by use of a
recently developed computational method.\cite{shin2018} It is based on
the concept that deformations in water's collective interfacial
molecular structure encode information about the details of
surface-water interactions.\cite{shin.water:2018} For this, the
probability distribution of molecular configurations, as characterized
by the three-dimensional vector, $\vec{\kappa} = (a,\cos{\theta_{\rm
    OH_1}},\cos{\theta_{\rm OH_2}})$. Here $a$ is the distance of the
oxygen atom position to the nearest point on the instantaneous water
interface, see Ref.\cite{willard_instantaneous_2010}, and $\theta_{\rm
  OH_1}$ and $\theta_{\rm OH_2}$ are the angles between the water OH
bonds and the interface normal.\\

\noindent
This method is used in the present work to compute the time dependent
quantity, $\delta \lambda_\mathrm{phob}^{(r)}(t)$, which describes the
local hydrophobicity (LH) of residue $r$, at time $t$. More
specifically, $\delta \lambda_\mathrm{phob}^{(r)}(t) =
\lambda_\mathrm{phob}^{(r)}(t) - \langle \lambda_\mathrm{phob}
\rangle_0$, where,
\begin{equation}
    \label{eqn:loglike}
    \lambda_\textrm{phob}^{(r)}(t) = -\frac{1}{{\sum_{a=1}^{N_a(r)}
        N_w(t;a)}}\sum_{a=1}^{N_a(r)} \sum_{i=1}^{N_w(t;a)} \ln{\left[
        \frac{P(\vec{\kappa}^{(i)}(t)|\textrm{phob})}{P(\vec{\kappa}^{(i)}(t)|\textrm{bulk})}
        \right]}.
\end{equation}
The sum over $N_a(r)$ includes the atoms in residue $r$, the summation
over $N_w(t;a)$ includes water molecules within a cut-off of 6 \AA\/
of atom $a$ at time $t$, and $\vec{\kappa}^{(i)}(t)$ denotes the
configuration of the $i$th molecule in this population. The
distribution $P(\vec{\kappa}\vert\mathrm{phob})$ is the probability to
find configuration $\vec{\kappa}$ at an ideal hydrophobic surface and
$P(\vec{\kappa} \vert \mathrm{bulk})$ is the probability to find that
same configuration in the isotropic environment of the bulk
liquid. These reference distributions were obtained by sampling the
orientational distribution of water at an ideal planar hydrophobic
silica surface and the bulk liquid, respectively, see
Ref.~\citenum{shin2018}. The quantity $\langle \lambda_\mathrm{phob}
\rangle_0$ is the equilibrium value of $\lambda_\mathrm{phob}$ for
configurational populations sampled from the ideal hydrophobic
reference system. Additional analyses with cutoffs of 5.5 \AA\/ and 7
\AA\/ were carried out and are reported in Figure \ref{sifig:plh}. The
shapes of $P({\rm LH})$ are largely independent of this choice, the
mode of the distributions can vary by up to 0.5 units (Thr41$\alpha$)
but are typically smaller than that, and the relative shifts of
$P({\rm LH})$ for the same residue in S1 and S2 are not affected by
the choice of the cutoff (see e.g. Thr41$\alpha$ in Figure
\ref{sifig:plh}).\\

\noindent
Values of $\delta \lambda_\textrm{phob}^{(r)}$ close to zero indicate
that water near residue $r$ exhibits orientations that are consistent with
those found at an ideal hydrophobic surface. Hydrophilic surfaces
interact with interfacial water molecules and lead to configurational
distributions that differ from those of an ideal hydrophobic surface.
These differences are typically reflected as values of $\delta
\lambda_\mathrm{phob}^{(r)} > 0$, with larger differences giving rise
to larger positive deviations in $\delta \lambda_\mathrm{phob}^{(r)}$.
Values of $\delta \lambda_\mathrm{phob}{(r)} \geq 0.5$ are indicative
of hydrophilicity. For the number of unique water configurations used
to compute $\delta \lambda_\mathrm{phob}^{(r)}$ here, fluctuations of
$\delta \lambda_\mathrm{phob}^{(r)}$ are expected to fall within
$-0.24 \leq \delta\lambda_\mathrm{phob}^{(r)} \leq 0.27$ (95\%
confidence interval) at the hydrophobic reference system. Therefore,
sustained values of $\lambda_\mathrm{phob}^{(r)} \geq 0.5$ are highly
indicative of local hydrophilicity. Fluctuations in $\delta
\lambda_\textrm{phob}^{(r)}$ as a function of time provide information
about changes in the local solvation environment.\\

\section*{Supplementary Material}
The supporting information reports figures for time series of LH$(t)$,
distributions $P({\rm LH})$, correlations between the modes of $P({\rm
  LH})$ for rigid and flexible T$_0$ and R$_4$, and radial
distribution functions between water and specific protein residues.

\section{Acknowledgment}
Support by the Swiss National Science Foundation through grants
200021-117810, the NCCR MUST (to MM), and the University of Basel is
acknowledged. The support of MK by the CHARMM Development Project is
gratefully acknowledged. The authors thank Marci Karplus and Victor
Ovchinnikov for help with editing the manuscripts.\\

\bibliography{article}

\providecommand{\latin}[1]{#1}
\makeatletter
\providecommand{\doi}
  {\begingroup\let\do\@makeother\dospecials
  \catcode`\{=1 \catcode`\}=2 \doi@aux}
\providecommand{\doi@aux}[1]{\endgroup\texttt{#1}}
\makeatother
\providecommand*\mcitethebibliography{\thebibliography}
\csname @ifundefined\endcsname{endmcitethebibliography}
  {\let\endmcitethebibliography\endthebibliography}{}
\begin{mcitethebibliography}{41}
\providecommand*\natexlab[1]{#1}
\providecommand*\mciteSetBstSublistMode[1]{}
\providecommand*\mciteSetBstMaxWidthForm[2]{}
\providecommand*\mciteBstWouldAddEndPuncttrue
  {\def\EndOfBibitem{\unskip.}}
\providecommand*\mciteBstWouldAddEndPunctfalse
  {\let\EndOfBibitem\relax}
\providecommand*\mciteSetBstMidEndSepPunct[3]{}
\providecommand*\mciteSetBstSublistLabelBeginEnd[3]{}
\providecommand*\EndOfBibitem{}
\mciteSetBstSublistMode{f}
\mciteSetBstMaxWidthForm{subitem}{(\alph{mcitesubitemcount})}
\mciteSetBstSublistLabelBeginEnd
  {\mcitemaxwidthsubitemform\space}
  {\relax}
  {\relax}

\bibitem[Teeter(1991)]{teeter:1991}
Teeter,~M.~M. Water-protein interactions: theory and experiment. \emph{Ann.
  Rev. Biophys. Biophys. Chem.} \textbf{1991}, \emph{20}, 577--600\relax
\mciteBstWouldAddEndPuncttrue
\mciteSetBstMidEndSepPunct{\mcitedefaultmidpunct}
{\mcitedefaultendpunct}{\mcitedefaultseppunct}\relax
\EndOfBibitem
\bibitem[Caronna \latin{et~al.}(2005)Caronna, Natali, and Cupane]{caronna:2005}
Caronna,~C.; Natali,~F.; Cupane,~A. Incoherent elastic and quasi-elastic
  neutron scattering investigation of hemoglobin dynamics. \emph{Biochem.}
  \textbf{2005}, \emph{116}, 219--225\relax
\mciteBstWouldAddEndPuncttrue
\mciteSetBstMidEndSepPunct{\mcitedefaultmidpunct}
{\mcitedefaultendpunct}{\mcitedefaultseppunct}\relax
\EndOfBibitem
\bibitem[Achterhold \latin{et~al.}(2002)Achterhold, Keppler, Ostermann,
  Van~B{\"u}rck, Sturhahn, Alp, and Parak]{parak:2002}
Achterhold,~K.; Keppler,~C.; Ostermann,~A.; Van~B{\"u}rck,~U.; Sturhahn,~W.;
  Alp,~E.; Parak,~F. Vibrational dynamics of myoglobin determined by the
  phonon-assisted M{\"o}ssbauer effect. \emph{Phys. Rep.} \textbf{2002},
  \emph{65}, 051916\relax
\mciteBstWouldAddEndPuncttrue
\mciteSetBstMidEndSepPunct{\mcitedefaultmidpunct}
{\mcitedefaultendpunct}{\mcitedefaultseppunct}\relax
\EndOfBibitem
\bibitem[Mouawad \latin{et~al.}(2005)Mouawad, Mar{\'e}chal, and
  Perahia]{mouawad:2005}
Mouawad,~L.; Mar{\'e}chal,~J.-D.; Perahia,~D. Internal cavities and ligand
  passageways in human hemoglobin characterized by molecular dynamics
  simulations. \emph{Biochimica et Biophysica Acta (BBA)-General Subjects}
  \textbf{2005}, \emph{1724}, 385--393\relax
\mciteBstWouldAddEndPuncttrue
\mciteSetBstMidEndSepPunct{\mcitedefaultmidpunct}
{\mcitedefaultendpunct}{\mcitedefaultseppunct}\relax
\EndOfBibitem
\bibitem[Fischer \latin{et~al.}({2011})Fischer, Olsen, Nam, and
  Karplus]{karplus.hb:2011}
Fischer,~S.; Olsen,~K.~W.; Nam,~K.; Karplus,~M. {Unsuspected pathway of the
  allosteric transition in hemoglobin}. \emph{Proc. Natl. Acad. Sci.}
  \textbf{{2011}}, \emph{{108}}, {5608--5613}\relax
\mciteBstWouldAddEndPuncttrue
\mciteSetBstMidEndSepPunct{\mcitedefaultmidpunct}
{\mcitedefaultendpunct}{\mcitedefaultseppunct}\relax
\EndOfBibitem
\bibitem[Yusuff \latin{et~al.}({2012})Yusuff, Babalola, Bussi, and
  Raugei]{yusuff:2012}
Yusuff,~O.~K.; Babalola,~J.~O.; Bussi,~G.; Raugei,~S. {Role of the Subunit
  Interactions in the Conformational Transitions in Adult Human Hemoglobin: An
  Explicit Solvent Molecular Dynamics Study}. \emph{J. Phys. Chem. B}
  \textbf{{2012}}, \emph{{116}}, {11004--11009}\relax
\mciteBstWouldAddEndPuncttrue
\mciteSetBstMidEndSepPunct{\mcitedefaultmidpunct}
{\mcitedefaultendpunct}{\mcitedefaultseppunct}\relax
\EndOfBibitem
\bibitem[Hub \latin{et~al.}(2010)Hub, Kubitzki, and de~Groot]{Hub_2010}
Hub,~J.~S.; Kubitzki,~M.~B.; de~Groot,~B.~L. Spontaneous Quaternary and
  Tertiary T-R Transitions of Human Hemoglobin in Molecular Dynamics
  Simulation. \emph{PLoS Comput Biol} \textbf{2010}, \emph{6}, e1000774\relax
\mciteBstWouldAddEndPuncttrue
\mciteSetBstMidEndSepPunct{\mcitedefaultmidpunct}
{\mcitedefaultendpunct}{\mcitedefaultseppunct}\relax
\EndOfBibitem
\bibitem[El~Hage \latin{et~al.}({2018})El~Hage, Hedin, Gupta, Meuwly, and
  Karplus]{MM.hb:2018}
El~Hage,~K.; Hedin,~F.; Gupta,~P.~K.; Meuwly,~M.; Karplus,~M. {Valid molecular
  dynamics simulations of human hemoglobin require a surprisingly large box
  size}. \emph{eLife} \textbf{{2018}}, \emph{{7}}, {e35560}\relax
\mciteBstWouldAddEndPuncttrue
\mciteSetBstMidEndSepPunct{\mcitedefaultmidpunct}
{\mcitedefaultendpunct}{\mcitedefaultseppunct}\relax
\EndOfBibitem
\bibitem[Pezzella \latin{et~al.}(2020)Pezzella, El~Hage, Niesen, Shin, Willard,
  Meuwly, and Karplus]{MM.hb:2020}
Pezzella,~M.; El~Hage,~K.; Niesen,~M.~J.; Shin,~S.; Willard,~A.~P.; Meuwly,~M.;
  Karplus,~M. Water dynamics around proteins: T-and R-states of hemoglobin and
  melittin. \emph{J. Phys. Chem. B} \textbf{2020}, \emph{124}, 6540--6554\relax
\mciteBstWouldAddEndPuncttrue
\mciteSetBstMidEndSepPunct{\mcitedefaultmidpunct}
{\mcitedefaultendpunct}{\mcitedefaultseppunct}\relax
\EndOfBibitem
\bibitem[Bernal(1965)]{bernal:1965}
Bernal,~J. The structure of water and its biological implications. Symposia of
  the Society for Experimental Biology. 1965; pp 17--32\relax
\mciteBstWouldAddEndPuncttrue
\mciteSetBstMidEndSepPunct{\mcitedefaultmidpunct}
{\mcitedefaultendpunct}{\mcitedefaultseppunct}\relax
\EndOfBibitem
\bibitem[Stadler \latin{et~al.}(2008)Stadler, Digel, Artmann, Embs, Zaccai, and
  B{\"u}ldt]{stadler:2008}
Stadler,~A.~M.; Digel,~I.; Artmann,~G.; Embs,~J.~P.; Zaccai,~G.; B{\"u}ldt,~G.
  Hemoglobin dynamics in red blood cells: correlation to body temperature.
  \emph{Biophys. J.} \textbf{2008}, \emph{95}, 5449--5461\relax
\mciteBstWouldAddEndPuncttrue
\mciteSetBstMidEndSepPunct{\mcitedefaultmidpunct}
{\mcitedefaultendpunct}{\mcitedefaultseppunct}\relax
\EndOfBibitem
\bibitem[Lesk \latin{et~al.}({1985})Lesk, Janin, Wodak, and
  Chothia]{chothia:1985}
Lesk,~A.; Janin,~J.; Wodak,~S.; Chothia,~C. {Hemoglobin - The surface buried
  between the alpha-1-beta-1 and alpha-2-beta-2 dimers in the deoxy and oxy
  structures}. \emph{J. Mol. Biol.} \textbf{{1985}}, \emph{{183}},
  {267--270}\relax
\mciteBstWouldAddEndPuncttrue
\mciteSetBstMidEndSepPunct{\mcitedefaultmidpunct}
{\mcitedefaultendpunct}{\mcitedefaultseppunct}\relax
\EndOfBibitem
\bibitem[Edelstein({1971})]{edelstein:1971}
Edelstein,~S. {Extensions of allosteric model for haemoglobin}. \emph{Nature}
  \textbf{{1971}}, \emph{{230}}, {224--227}\relax
\mciteBstWouldAddEndPuncttrue
\mciteSetBstMidEndSepPunct{\mcitedefaultmidpunct}
{\mcitedefaultendpunct}{\mcitedefaultseppunct}\relax
\EndOfBibitem
\bibitem[Kister \latin{et~al.}(1987)Kister, Poyart, and Edelstein]{kister:1987}
Kister,~J.; Poyart,~C.; Edelstein,~S.~J. An expanded two-state allosteric model
  for interactions of human hemoglobin A with nonsaturating concentrations of
  2, 3-diphosphoglycerate. \emph{J. Biol. Chem.} \textbf{1987}, \emph{262},
  12085--12091\relax
\mciteBstWouldAddEndPuncttrue
\mciteSetBstMidEndSepPunct{\mcitedefaultmidpunct}
{\mcitedefaultendpunct}{\mcitedefaultseppunct}\relax
\EndOfBibitem
\bibitem[Chothia \latin{et~al.}({1976})Chothia, Wodak, and Janin]{janin:1976}
Chothia,~C.; Wodak,~S.; Janin,~J. {Role of Subunit Interfaces in Allosteric
  Mechanism of Hemoglobin}. \emph{Proc. Natl. Acad. Sci.} \textbf{{1976}},
  \emph{{73}}, {3793--3797}\relax
\mciteBstWouldAddEndPuncttrue
\mciteSetBstMidEndSepPunct{\mcitedefaultmidpunct}
{\mcitedefaultendpunct}{\mcitedefaultseppunct}\relax
\EndOfBibitem
\bibitem[Rossky \latin{et~al.}({1979})Rossky, Karplus, and Rahman]{Rossky:1979}
Rossky,~P.; Karplus,~M.; Rahman,~A. {Model for the simulation of an aqueous
  dipeptide solution}. \emph{{Biopolymers}} \textbf{{1979}}, \emph{{18}},
  {825--854}\relax
\mciteBstWouldAddEndPuncttrue
\mciteSetBstMidEndSepPunct{\mcitedefaultmidpunct}
{\mcitedefaultendpunct}{\mcitedefaultseppunct}\relax
\EndOfBibitem
\bibitem[Cheng and Rossky({1998})Cheng, and Rossky]{rossky:1998}
Cheng,~Y.; Rossky,~P. {Surface topography dependence of biomolecular
  hydrophobic hydration}. \emph{Nature} \textbf{{1998}}, \emph{{392}},
  {696--699}\relax
\mciteBstWouldAddEndPuncttrue
\mciteSetBstMidEndSepPunct{\mcitedefaultmidpunct}
{\mcitedefaultendpunct}{\mcitedefaultseppunct}\relax
\EndOfBibitem
\bibitem[Chandler({2005})]{chandler:2005}
Chandler,~D. {Interfaces and the driving force of hydrophobic assembly}.
  \emph{Nature} \textbf{{2005}}, \emph{{437}}, {640--647}\relax
\mciteBstWouldAddEndPuncttrue
\mciteSetBstMidEndSepPunct{\mcitedefaultmidpunct}
{\mcitedefaultendpunct}{\mcitedefaultseppunct}\relax
\EndOfBibitem
\bibitem[Chandler and Varilly({2012})Chandler, and
  Varilly]{chandler.varenna:2012}
Chandler,~D.; Varilly,~P. {Lectures on molecular- and nano-scale fluctuations
  in water}. {Complex Materials in Physics and Biology}. {2012}; pp {75--111},
  {176th Course of the International School of Physics Enrico Fermi on Complex
  Materials in Physics and Biology, Varenna, ITALY, JUN 29-JUL 09, 2010}\relax
\mciteBstWouldAddEndPuncttrue
\mciteSetBstMidEndSepPunct{\mcitedefaultmidpunct}
{\mcitedefaultendpunct}{\mcitedefaultseppunct}\relax
\EndOfBibitem
\bibitem[Ladbury(1996)]{ladbury:1996}
Ladbury,~J.~E. Just add water! The effect of water on the specificity of
  protein-ligand binding sites and its potential application to drug design.
  \emph{Chem. Biol.} \textbf{1996}, \emph{3}, 973--980\relax
\mciteBstWouldAddEndPuncttrue
\mciteSetBstMidEndSepPunct{\mcitedefaultmidpunct}
{\mcitedefaultendpunct}{\mcitedefaultseppunct}\relax
\EndOfBibitem
\bibitem[Mahmoud \latin{et~al.}(2020)Mahmoud, Masters, Yang, and
  Lill]{lill:2020}
Mahmoud,~A.~H.; Masters,~M.~R.; Yang,~Y.; Lill,~M.~A. Elucidating the multiple
  roles of hydration for accurate protein-ligand binding prediction via deep
  learning. \emph{Comm. Chem.} \textbf{2020}, \emph{3}, 1--13\relax
\mciteBstWouldAddEndPuncttrue
\mciteSetBstMidEndSepPunct{\mcitedefaultmidpunct}
{\mcitedefaultendpunct}{\mcitedefaultseppunct}\relax
\EndOfBibitem
\bibitem[Rizzi \latin{et~al.}(2021)Rizzi, Bonati, Ansari, and
  Parrinello]{rizzi:2021}
Rizzi,~V.; Bonati,~L.; Ansari,~N.; Parrinello,~M. The role of water in
  host-guest interaction. \emph{Nat. Comm.} \textbf{2021}, \emph{12},
  1--7\relax
\mciteBstWouldAddEndPuncttrue
\mciteSetBstMidEndSepPunct{\mcitedefaultmidpunct}
{\mcitedefaultendpunct}{\mcitedefaultseppunct}\relax
\EndOfBibitem
\bibitem[Chothia(1974)]{chothia:1974}
Chothia,~C. Hydrophobic bonding and accessible surface area in proteins.
  \emph{Nature} \textbf{1974}, \emph{248}, 338--339\relax
\mciteBstWouldAddEndPuncttrue
\mciteSetBstMidEndSepPunct{\mcitedefaultmidpunct}
{\mcitedefaultendpunct}{\mcitedefaultseppunct}\relax
\EndOfBibitem
\bibitem[Zhou \latin{et~al.}(2003)Zhou, Zhou, and Karplus]{zhou:2003}
Zhou,~Y.; Zhou,~H.; Karplus,~M. Cooperativity in Scapharca dimeric hemoglobin:
  simulation of binding intermediates and elucidation of the role of
  interfacial water. \emph{J. Mol. Biol.} \textbf{2003}, \emph{326},
  593--606\relax
\mciteBstWouldAddEndPuncttrue
\mciteSetBstMidEndSepPunct{\mcitedefaultmidpunct}
{\mcitedefaultendpunct}{\mcitedefaultseppunct}\relax
\EndOfBibitem
\bibitem[Gupta and Meuwly(2016)Gupta, and Meuwly]{MM.hb:2016}
Gupta,~P.~K.; Meuwly,~M. Ligand and interfacial dynamics in a homodimeric
  hemoglobin. \emph{Struc. Dyn.} \textbf{2016}, \emph{3}, 012003\relax
\mciteBstWouldAddEndPuncttrue
\mciteSetBstMidEndSepPunct{\mcitedefaultmidpunct}
{\mcitedefaultendpunct}{\mcitedefaultseppunct}\relax
\EndOfBibitem
\bibitem[Colombo \latin{et~al.}(1992)Colombo, Rau, and Parsegian]{colombo:1992}
Colombo,~M.~F.; Rau,~D.~C.; Parsegian,~V.~A. Protein solvation in allosteric
  regulation: a water effect on hemoglobin. \emph{Science} \textbf{1992},
  \emph{256}, 655--659\relax
\mciteBstWouldAddEndPuncttrue
\mciteSetBstMidEndSepPunct{\mcitedefaultmidpunct}
{\mcitedefaultendpunct}{\mcitedefaultseppunct}\relax
\EndOfBibitem
\bibitem[Monod and Jacob(1961)Monod, and Jacob]{monod:1961}
Monod,~J.; Jacob,~F. General conclusions: teleonomic mechanisms in cellular
  metabolism, growth, and differentiation. Cold Spring Harbor symposia on
  quantitative biology. 1961; pp 389--401\relax
\mciteBstWouldAddEndPuncttrue
\mciteSetBstMidEndSepPunct{\mcitedefaultmidpunct}
{\mcitedefaultendpunct}{\mcitedefaultseppunct}\relax
\EndOfBibitem
\bibitem[Monod \latin{et~al.}(1965)Monod, Wyman, and Changeux]{monod:1965}
Monod,~J.; Wyman,~J.; Changeux,~J.-P. On the nature of allosteric transitions:
  a plausible model. \emph{J. Mol. Biol.} \textbf{1965}, \emph{12},
  88--118\relax
\mciteBstWouldAddEndPuncttrue
\mciteSetBstMidEndSepPunct{\mcitedefaultmidpunct}
{\mcitedefaultendpunct}{\mcitedefaultseppunct}\relax
\EndOfBibitem
\bibitem[Pauling(1935)]{pauling:1935}
Pauling,~L. The oxygen equilibrium of hemoglobin and its structural
  interpretation. \emph{Proc. Natl. Acad. Sci.} \textbf{1935}, \emph{21},
  186--191\relax
\mciteBstWouldAddEndPuncttrue
\mciteSetBstMidEndSepPunct{\mcitedefaultmidpunct}
{\mcitedefaultendpunct}{\mcitedefaultseppunct}\relax
\EndOfBibitem
\bibitem[Koshland~Jr \latin{et~al.}(1966)Koshland~Jr, N{\'e}methy, and
  Filmer]{koshland:1966}
Koshland~Jr,~D.~E.; N{\'e}methy,~G.; Filmer,~D. Comparison of experimental
  binding data and theoretical models in proteins containing subunits.
  \emph{Biochem.} \textbf{1966}, \emph{5}, 365--385\relax
\mciteBstWouldAddEndPuncttrue
\mciteSetBstMidEndSepPunct{\mcitedefaultmidpunct}
{\mcitedefaultendpunct}{\mcitedefaultseppunct}\relax
\EndOfBibitem
\bibitem[Szabo and Karplus({1972})Szabo, and Karplus]{szabo:1972}
Szabo,~A.; Karplus,~M. {Mathematical model for structure-function relations in
  hemoglobin}. \emph{J. Mol. Biol.} \textbf{{1972}}, \emph{{72}},
  {163--197}\relax
\mciteBstWouldAddEndPuncttrue
\mciteSetBstMidEndSepPunct{\mcitedefaultmidpunct}
{\mcitedefaultendpunct}{\mcitedefaultseppunct}\relax
\EndOfBibitem
\bibitem[Klauda \latin{et~al.}(2010)Klauda, Venable, Freites, O’Connor,
  Tobias, Mondragon-Ramirez, Vorobyov, MacKerell, and Pastor]{charmm36:2010}
Klauda,~J.~B.; Venable,~R.~M.; Freites,~J.~A.; O’Connor,~J.~W.;
  Tobias,~D.~J.; Mondragon-Ramirez,~C.; Vorobyov,~I.; MacKerell,~A.~D.;
  Pastor,~R.~W. Update of the CHARMM All-Atom Additive Force Field for Lipids:
  Validation on Six Lipid Types. \emph{J. Phys. Chem. B} \textbf{2010},
  \emph{114}, 7830--7843\relax
\mciteBstWouldAddEndPuncttrue
\mciteSetBstMidEndSepPunct{\mcitedefaultmidpunct}
{\mcitedefaultendpunct}{\mcitedefaultseppunct}\relax
\EndOfBibitem
\bibitem[Jorgensen \latin{et~al.}(1983)Jorgensen, Chandrasekhar, Madura, Impey,
  and Klein]{jorgensen-tip3p}
Jorgensen,~W.~L.; Chandrasekhar,~J.; Madura,~J.~D.; Impey,~R.~W.; Klein,~M.~L.
  Comparison of simple potential functions for simulating liquid water.
  \emph{J. Chem. Phys.} \textbf{1983}, \emph{79}, 926--935\relax
\mciteBstWouldAddEndPuncttrue
\mciteSetBstMidEndSepPunct{\mcitedefaultmidpunct}
{\mcitedefaultendpunct}{\mcitedefaultseppunct}\relax
\EndOfBibitem
\bibitem[Park \latin{et~al.}({2006})Park, Yokoyama, Shibayama, Shiro, and
  Tame]{tame:2006}
Park,~S.-Y.; Yokoyama,~T.; Shibayama,~N.; Shiro,~Y.; Tame,~J. R.~H. {1.25
  angstrom resolution crystal structures of human haemoglobin in the oxy, deoxy
  and carbonmonoxy forms}. \emph{J. Mol. Biol.} \textbf{{2006}}, \emph{{360}},
  {690--701}\relax
\mciteBstWouldAddEndPuncttrue
\mciteSetBstMidEndSepPunct{\mcitedefaultmidpunct}
{\mcitedefaultendpunct}{\mcitedefaultseppunct}\relax
\EndOfBibitem
\bibitem[Eastman \latin{et~al.}({2017})Eastman, Swails, Chodera, McGibbon,
  Zhao, Beauchamp, Wang, Simmonett, Harrigan, Stern, Wiewiora, Brooks, and
  Pande]{openMM}
Eastman,~P.; Swails,~J.; Chodera,~J.~D.; McGibbon,~R.~T.; Zhao,~Y.;
  Beauchamp,~K.~A.; Wang,~L.-P.; Simmonett,~A.~C.; Harrigan,~M.~P.;
  Stern,~C.~D.; Wiewiora,~R.~P.; Brooks,~B.~R.; Pande,~V.~S. {OpenMM 7: Rapid
  development of high performance algorithms for molecular dynamics}.
  \emph{{PLOS Comp. Biol.}} \textbf{{2017}}, \emph{{13}}\relax
\mciteBstWouldAddEndPuncttrue
\mciteSetBstMidEndSepPunct{\mcitedefaultmidpunct}
{\mcitedefaultendpunct}{\mcitedefaultseppunct}\relax
\EndOfBibitem
\bibitem[Buck \latin{et~al.}(2006)Buck, Bouguet-Bonnet, Pastor, and Alexander
  D.~MacKerell]{cmap06}
Buck,~M.; Bouguet-Bonnet,~S.; Pastor,~R.~W.; Alexander D.~MacKerell,~J.
  Importance of the CMAP Correction to the CHARMM22 Protein Force Field:
  Dynamics of Hen Lysozyme. \emph{Biophys J.} \textbf{2006}, \emph{90},
  L36--l38\relax
\mciteBstWouldAddEndPuncttrue
\mciteSetBstMidEndSepPunct{\mcitedefaultmidpunct}
{\mcitedefaultendpunct}{\mcitedefaultseppunct}\relax
\EndOfBibitem
\bibitem[Darden \latin{et~al.}(1993)Darden, York, and Pedersen]{Darden1993}
Darden,~T.; York,~D.; Pedersen,~L. Particle Mesh Ewald: An Nlog(N) Method for
  Ewald Sums in Large Systems. \emph{J. Chem. Phys.} \textbf{1993}, \emph{98},
  10089--10092\relax
\mciteBstWouldAddEndPuncttrue
\mciteSetBstMidEndSepPunct{\mcitedefaultmidpunct}
{\mcitedefaultendpunct}{\mcitedefaultseppunct}\relax
\EndOfBibitem
\bibitem[Shin and Willard(2018)Shin, and Willard]{shin2018}
Shin,~S.; Willard,~A.~P. Characterizing Hydration Properties Based on the
  Orientational Structure of Interfacial Water Molecules. \emph{J. Chem. Theor.
  Comput.} \textbf{2018}, \emph{14}, 461--465\relax
\mciteBstWouldAddEndPuncttrue
\mciteSetBstMidEndSepPunct{\mcitedefaultmidpunct}
{\mcitedefaultendpunct}{\mcitedefaultseppunct}\relax
\EndOfBibitem
\bibitem[Shin and Willard({2018})Shin, and Willard]{shin.water:2018}
Shin,~S.; Willard,~A.~P. {Water's Interfacial Hydrogen Bonding Structure
  Reveals the Effective Strength of Surface-Water Interactions}. \emph{J. Phys.
  Chem. B} \textbf{{2018}}, \emph{{122}}, {6781--6789}\relax
\mciteBstWouldAddEndPuncttrue
\mciteSetBstMidEndSepPunct{\mcitedefaultmidpunct}
{\mcitedefaultendpunct}{\mcitedefaultseppunct}\relax
\EndOfBibitem
\bibitem[Willard and Chandler(2010)Willard, and
  Chandler]{willard_instantaneous_2010}
Willard,~A.~P.; Chandler,~D. Instantaneous {Liquid} {Interfaces}. \emph{J.
  Phys. Chem. B} \textbf{2010}, \emph{114}, 1954--1958\relax
\mciteBstWouldAddEndPuncttrue
\mciteSetBstMidEndSepPunct{\mcitedefaultmidpunct}
{\mcitedefaultendpunct}{\mcitedefaultseppunct}\relax
\EndOfBibitem
\end{mcitethebibliography}

\clearpage

\renewcommand{\thepage}{S\arabic{page}}
\renewcommand{\thetable}{S\arabic{table}}
\renewcommand{\thefigure}{S\arabic{figure}}
\renewcommand{\theequation}{S\arabic{equation}}
\renewcommand{\thesection}{S\arabic{section}}
\setcounter{figure}{0}  
\setcounter{section}{0}
\setcounter{page}{1}

\noindent
{\bf Supporting Information: Water Dynamics around T0 vs. R4 from
  Local Hydrophobicity Analysis}

\begin{figure}
\centering
\includegraphics[scale=0.6]{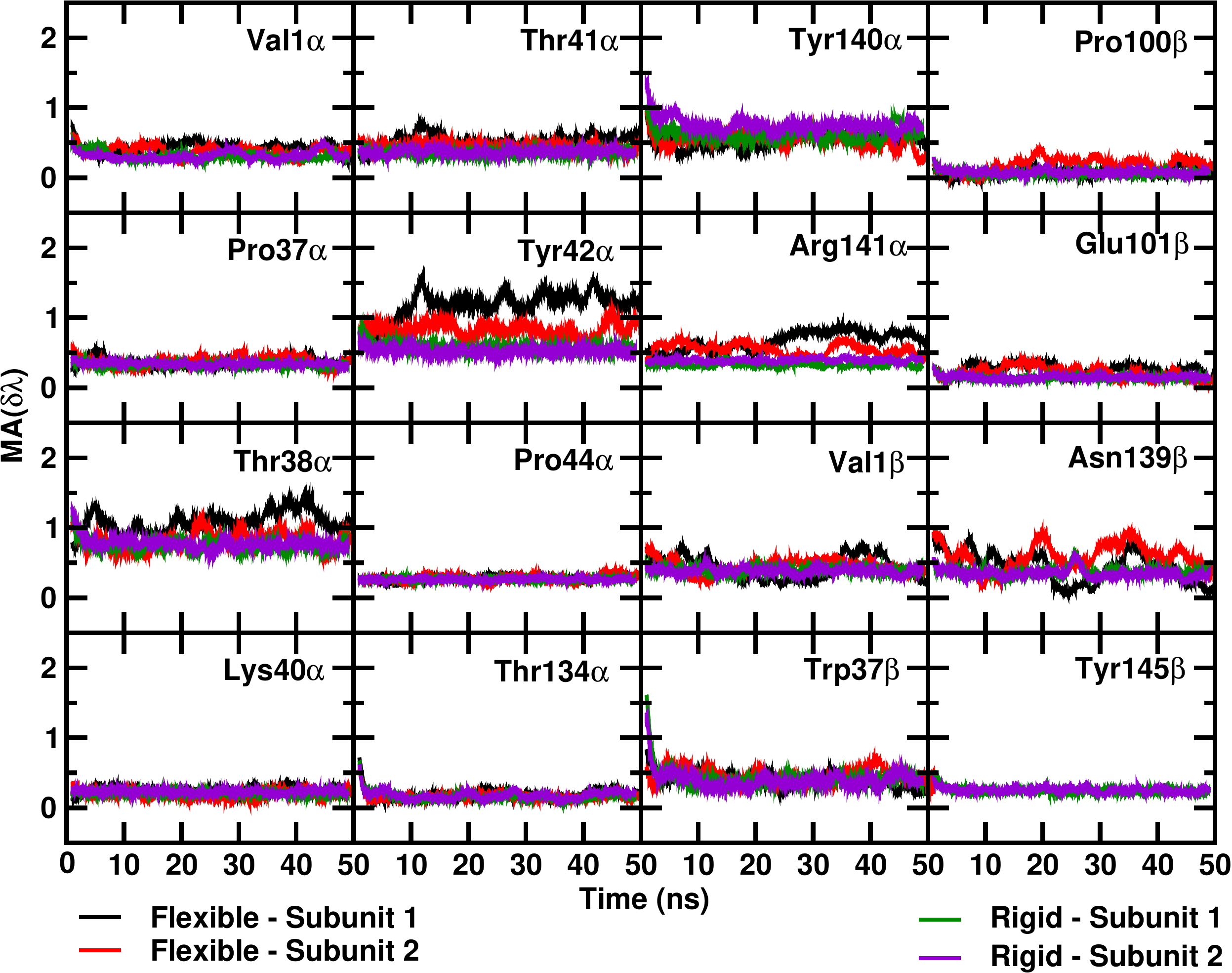} 
\caption{LH time series for flexible and rigid R$_4$ Hb tetramer from
  50 ns simulations: Time series for the two monomers S1 and S2 are
  represented in black ($\alpha_1\beta_1$) and red ($\alpha_2\beta_2$)
  for flexible and in green ($\alpha_1\beta_1$) and violet
  ($\alpha_2\beta_2$) for rigid tetramer. A window of 200 points was
  used for the running average.}
\label{sifig:lh.r4}
\end{figure}

\begin{figure}
\centering
\includegraphics[scale=0.6]{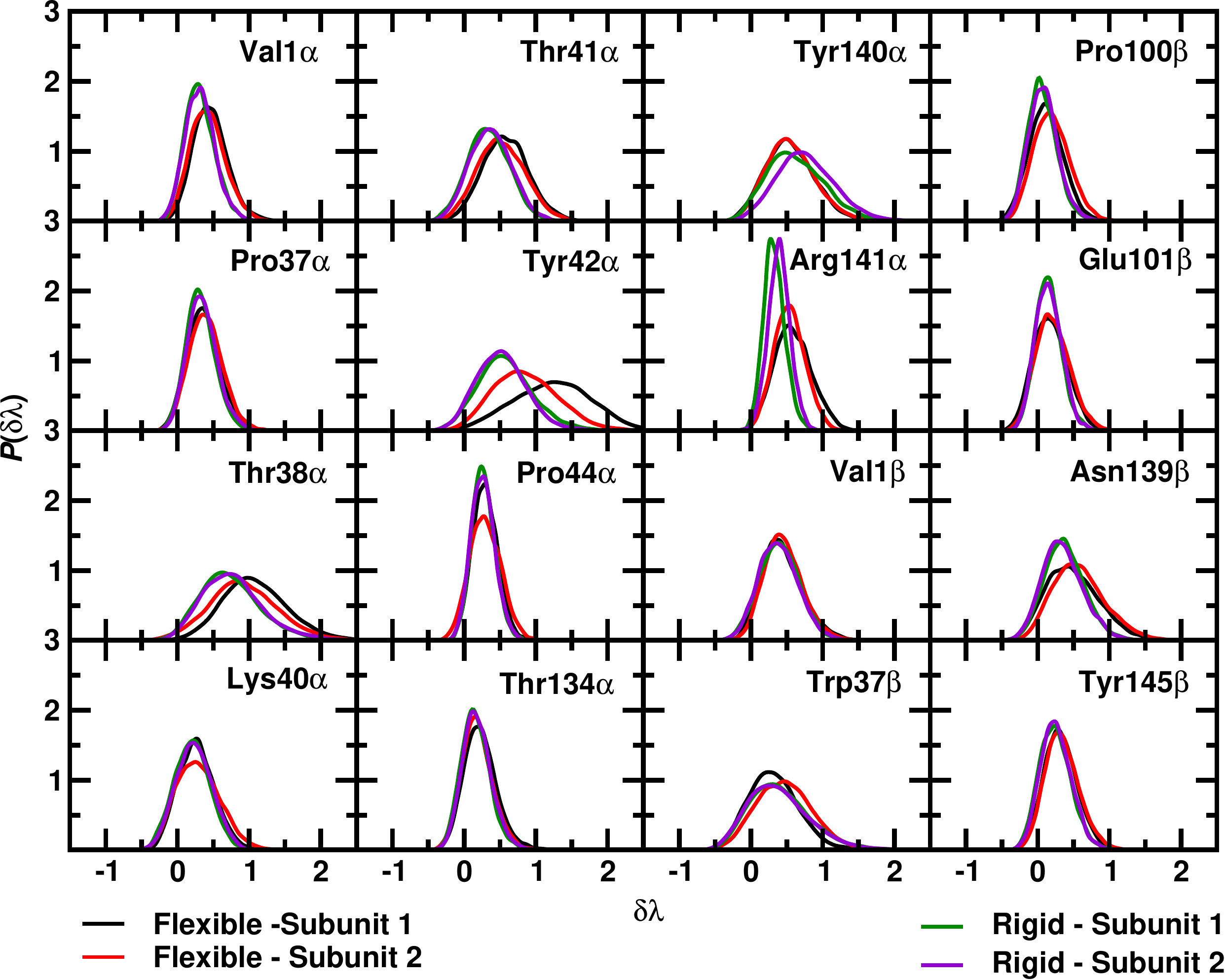} 
\caption{$P({\rm LH})$ for flexible and rigid R$_4$ tetramer from 50
  ns simulations. The two different monomers are represented in black
  ($\alpha_1\beta_1$) and red ($\alpha_2\beta_2$) for flexible and in
  green ($\alpha_1\beta_1$) and violet ($\alpha_2\beta_2$) for rigid
  tetramer.}
\label{sifig:plh.r4}
\end{figure}

\begin{figure}
\centering \includegraphics[scale=0.7]{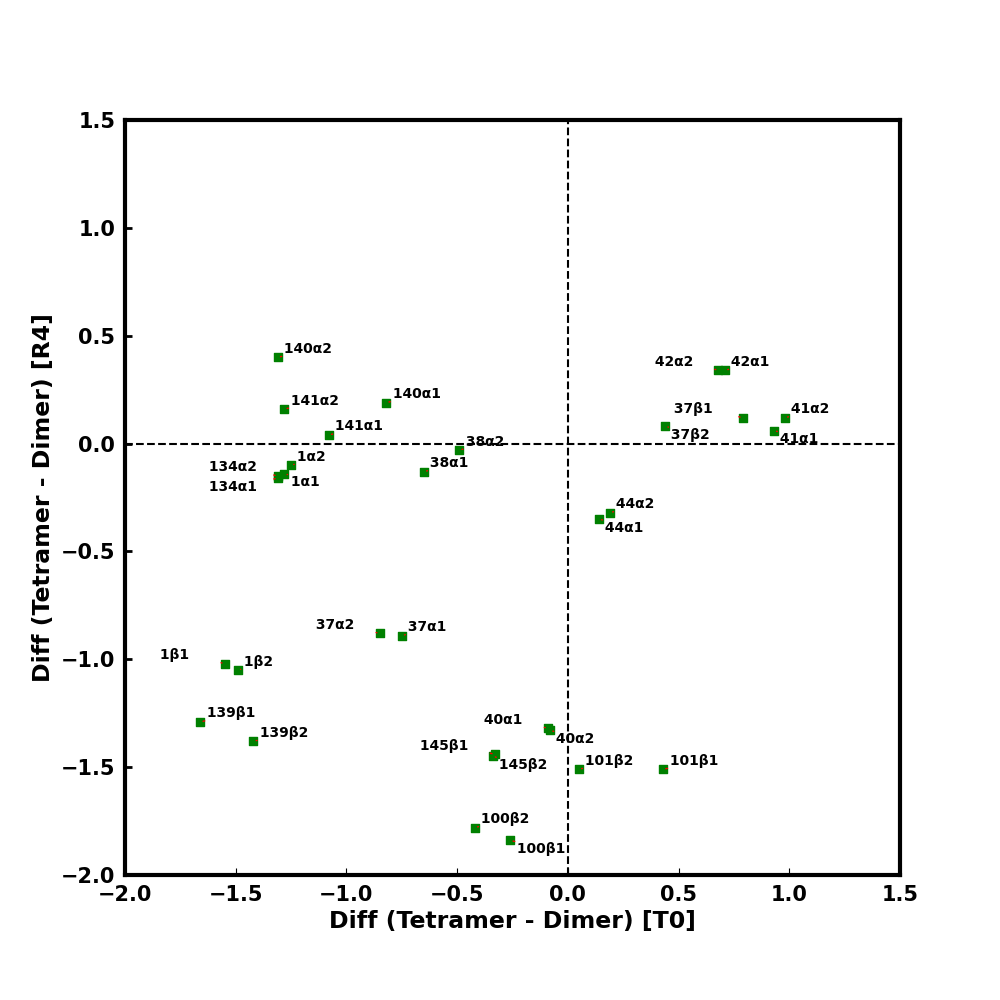}
\caption{Difference in max$P({\rm LH})$ of (tetramer - dimer) for
  T$_0$ vs. R$_4$ from rigid simulations.}
\label{sifig:corr_diff_rig}
\end{figure}

\begin{figure}
\centering \includegraphics[scale=0.7]{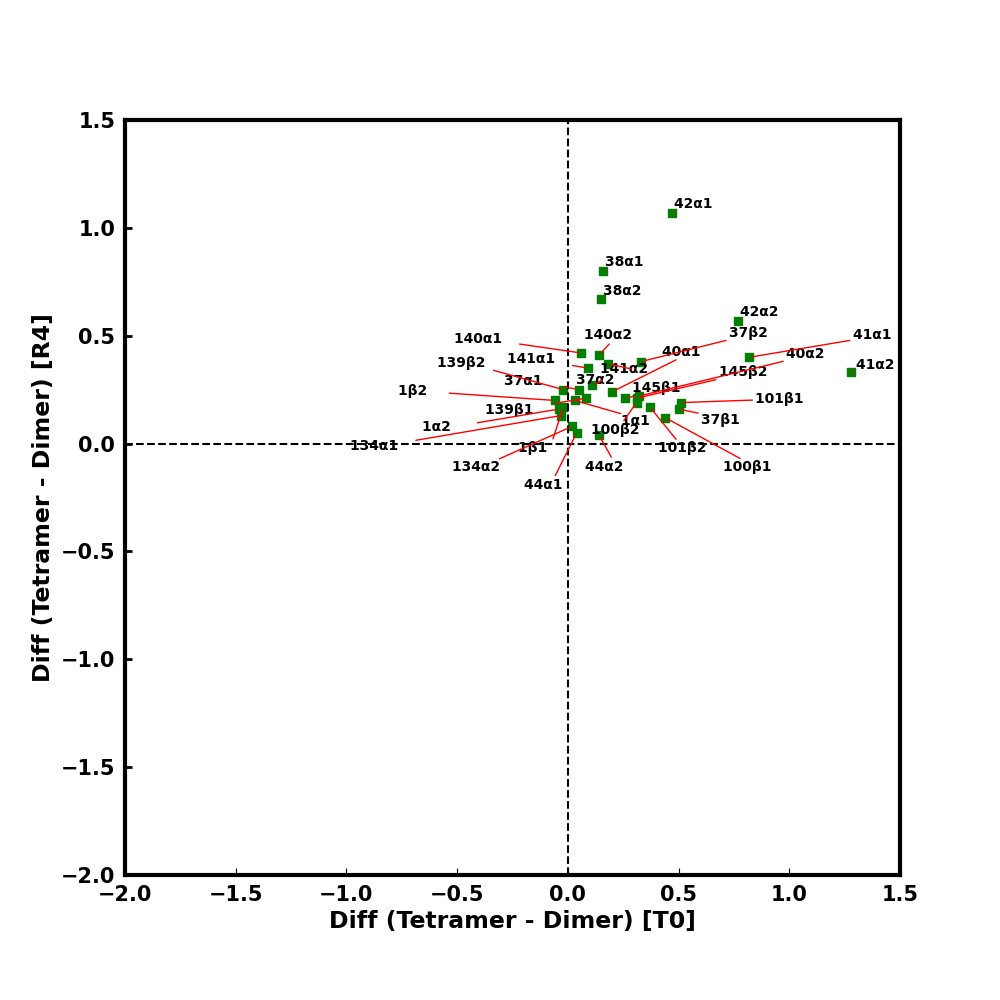}
\caption{Difference in max$P({\rm LH})$ of (tetramer - dimer) for
  T$_0$ vs. R$_4$ from flexible simulations.}
\label{sifig:corr_diff_flex}
\end{figure}

\begin{figure}
\centering \includegraphics[scale=0.5]{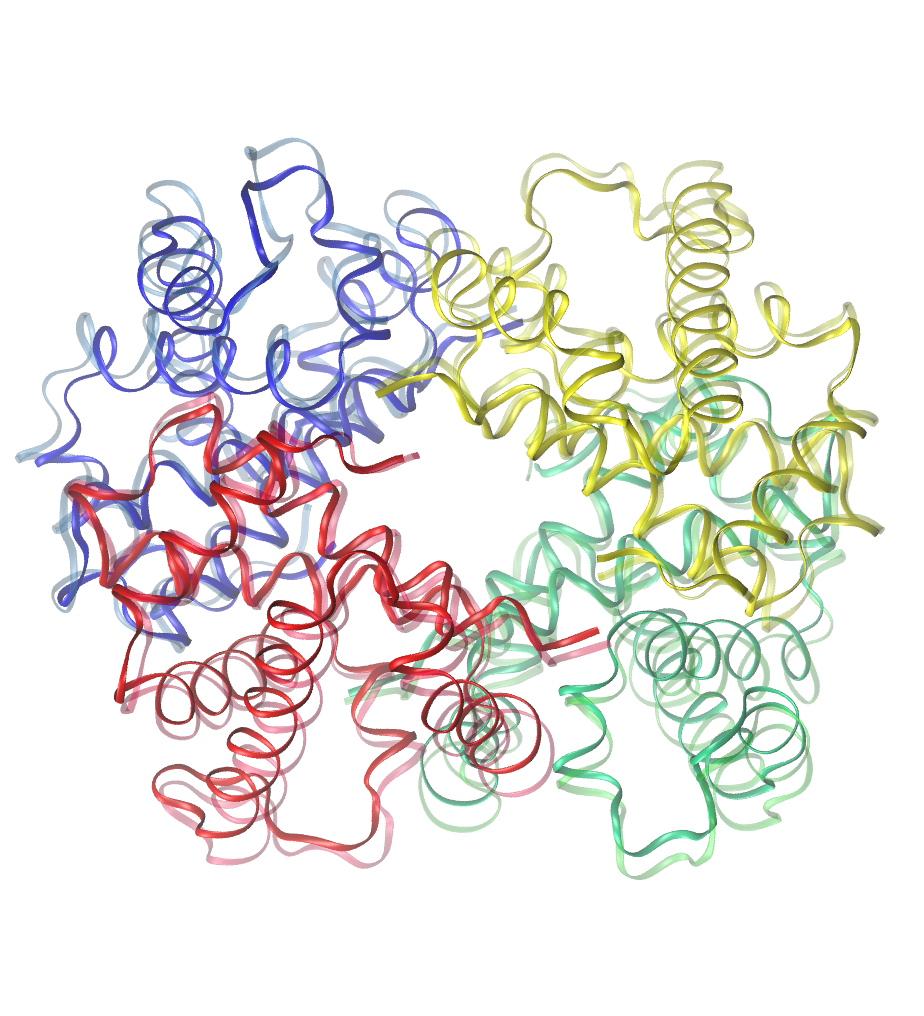}
\caption{Superposition of the rigid (semi transparent) and dynamically
  averaged (full color) structures for T$_0$. The color code and
  orientation is that of Figure 1.}
\label{sifig:rmsd-t0}
\end{figure}

\begin{figure}
\centering \includegraphics[scale=0.5]{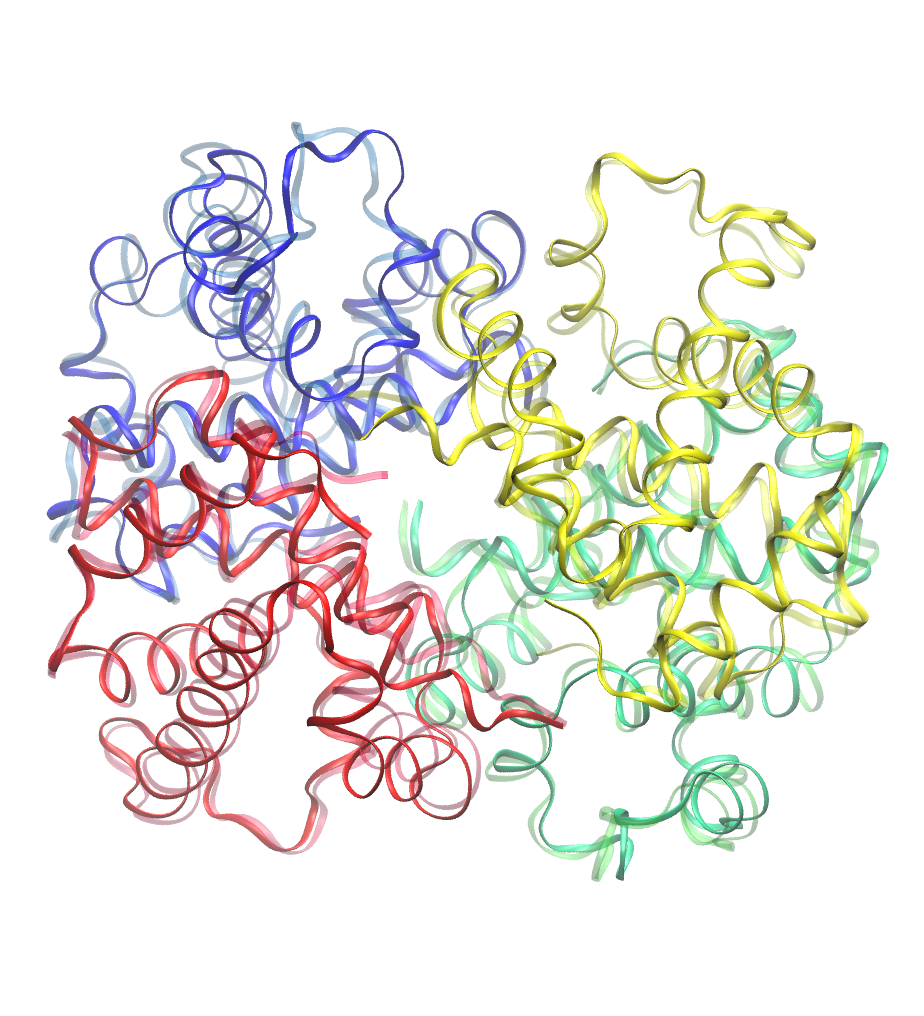}
\caption{Superposition of the rigid (semi transparent) and dynamically
  averaged (full color) structures for R$_4$. The color code and
  orientation is that of Figure 1.}
\label{sifig:rmsd-r4}
\end{figure}

\begin{figure}
  \centering \includegraphics[scale=0.7]{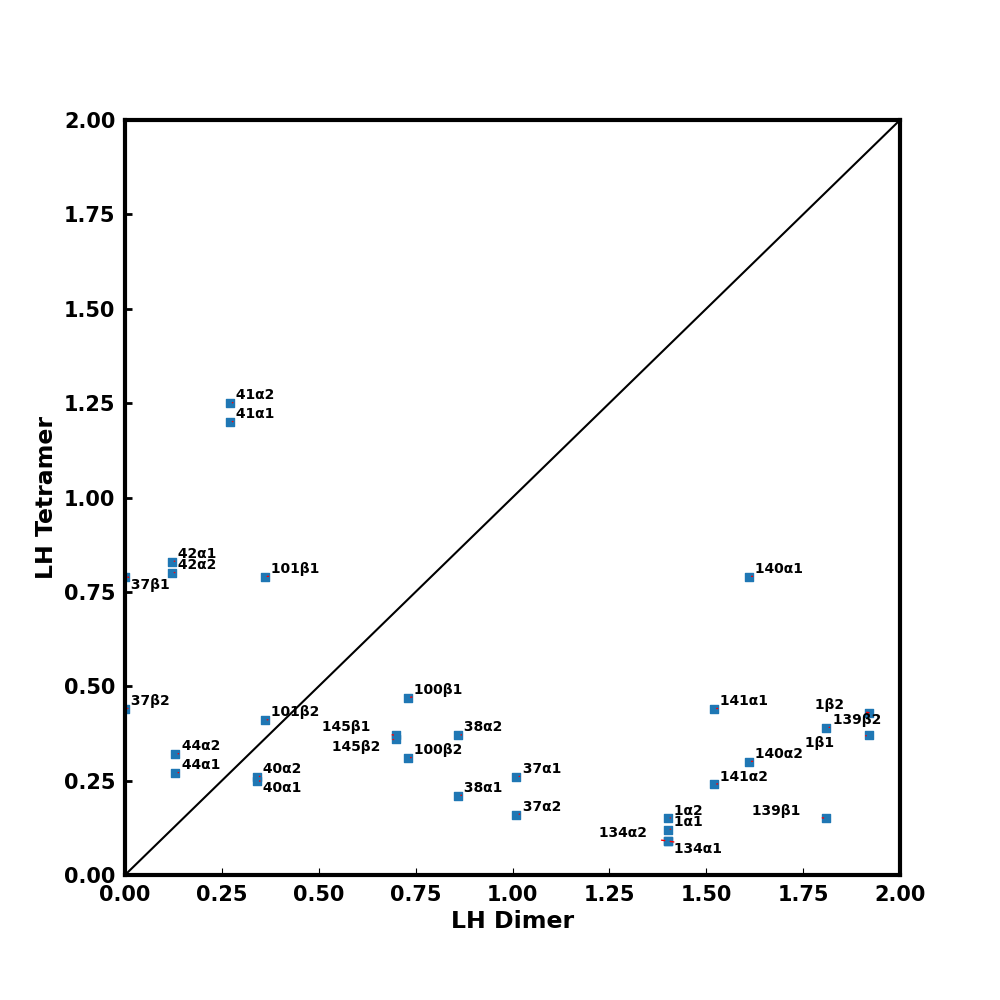}
\caption{Comparison of max$P({\rm LH})$ for dimer vs. tetramer in
  rigid T$_0$ tetramer.}
\label{sifig:corr_t0_rigid}
\end{figure}

\begin{figure}
  \centering \includegraphics[scale=0.7]{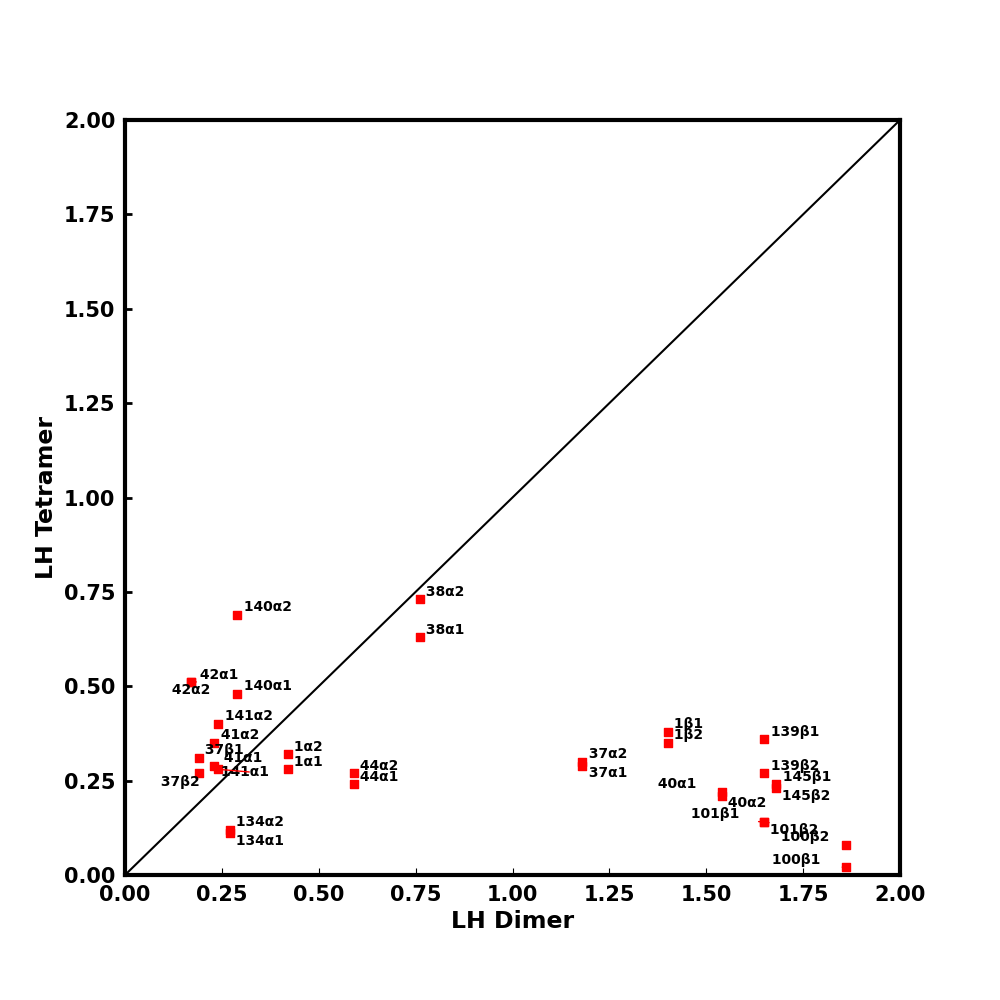}
\caption{Comparison of 
max$P({\rm LH})$ for dimer vs. tetramer in rigid R$_4$ tetramer.}
\label{sifig:corr_r4_rigid}
\end{figure}

\begin{figure}
\centering \includegraphics[scale=0.7]{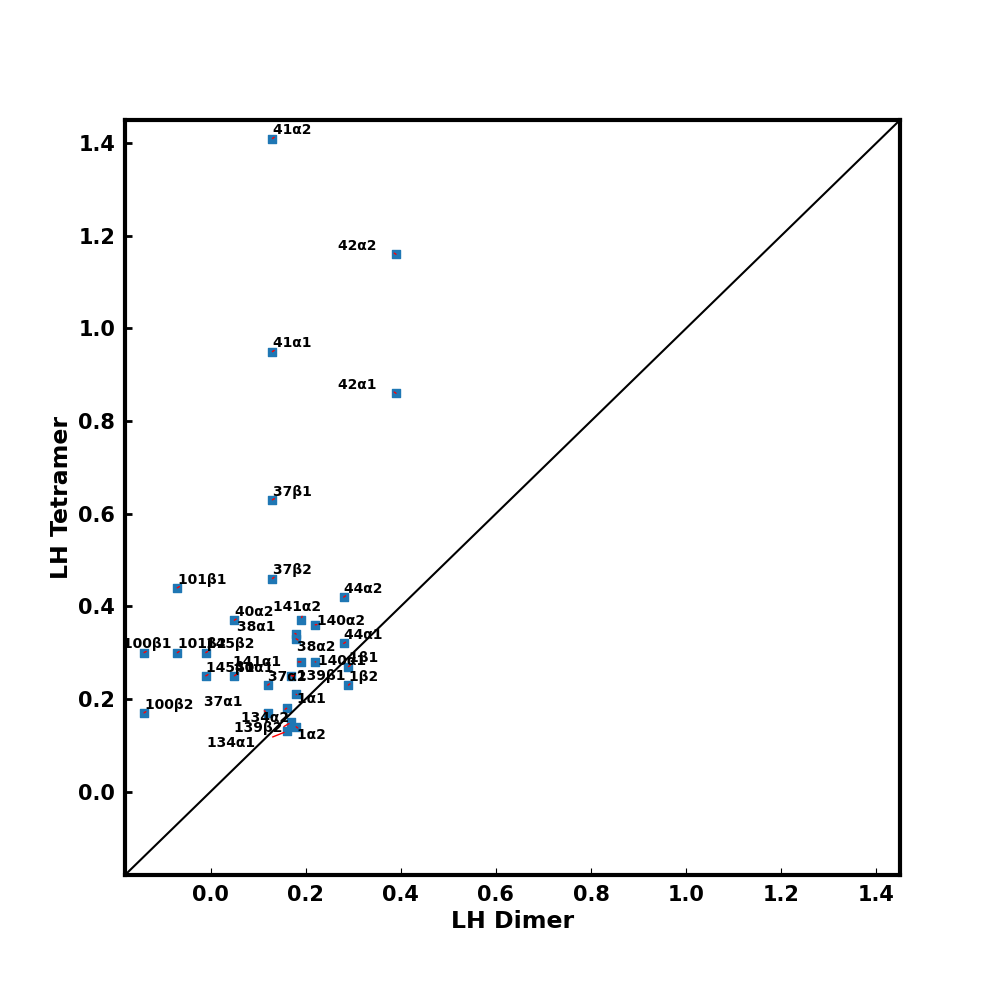}
\caption{Comparison of 
max$P({\rm LH})$ for dimer vs. tetramer in
  flexible T$_0$ tetramer.}
\label{sifig:corr_t0_flex}
\end{figure}

\begin{figure}
\centering \includegraphics[scale=0.7]{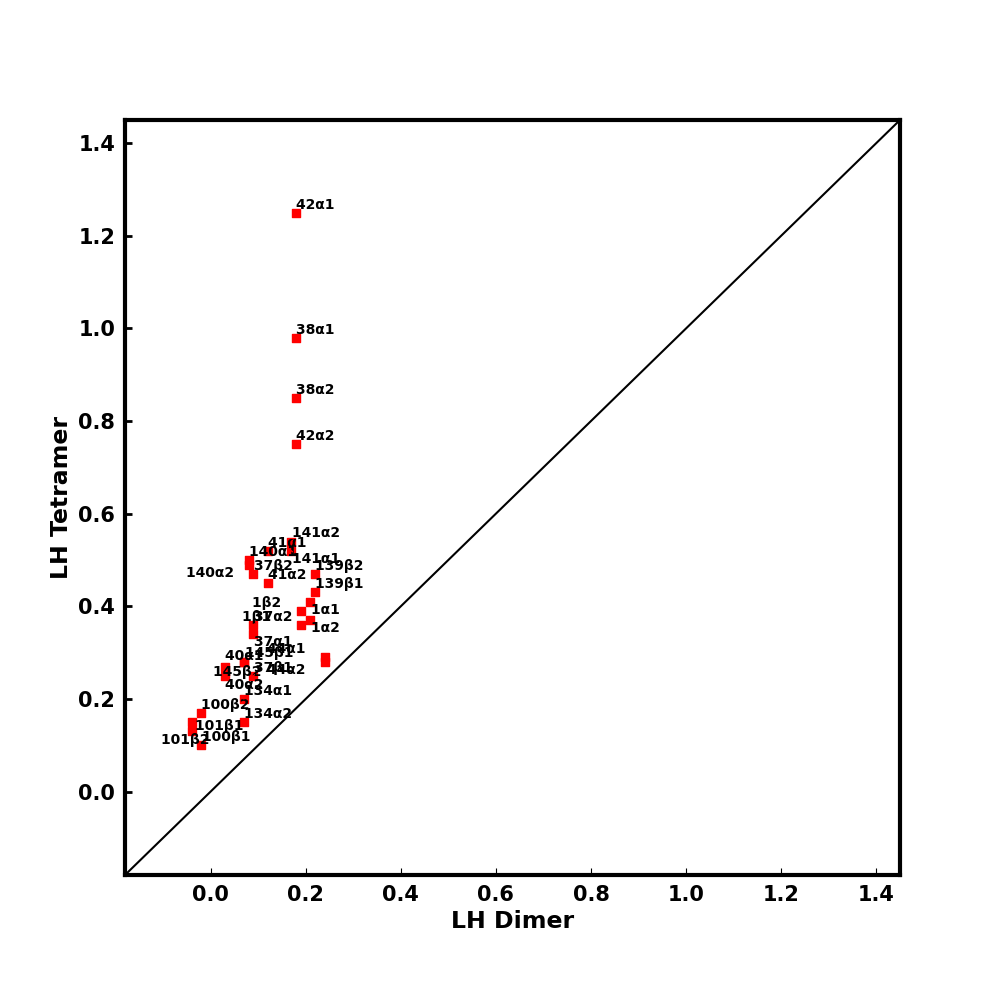}
\caption{Comparison of 
max$P({\rm LH})$ for dimer vs. tetramer in
  flexible R$_4$ tetramer.}
\label{sifig:corr_r4_flex}
\end{figure}

\begin{figure}
\centering \includegraphics[scale=0.7]{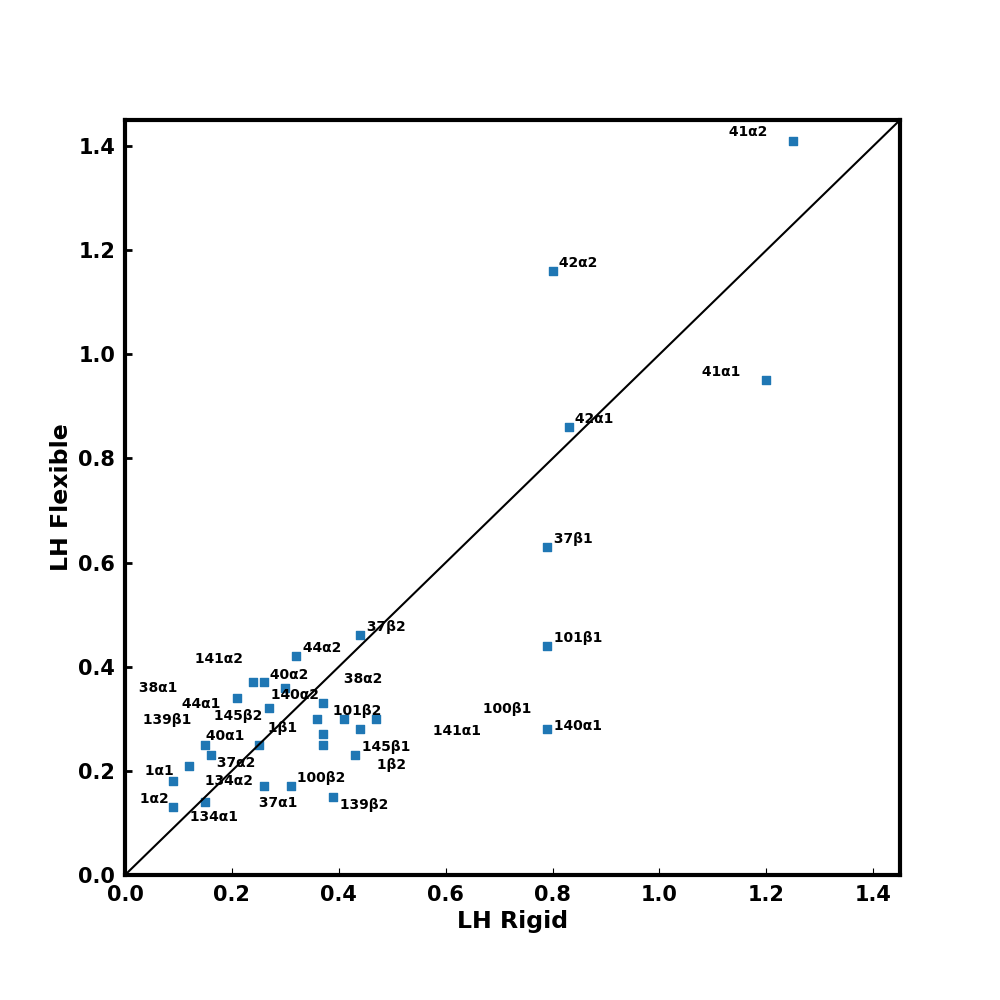}
\caption{Comparison of max$P({\rm LH})$ for rigid vs. flexible T$_0$ tetramer.}
\label{sifig:corr_rigid_flex_t0}
\end{figure}

\begin{figure}
\centering \includegraphics[scale=0.7]{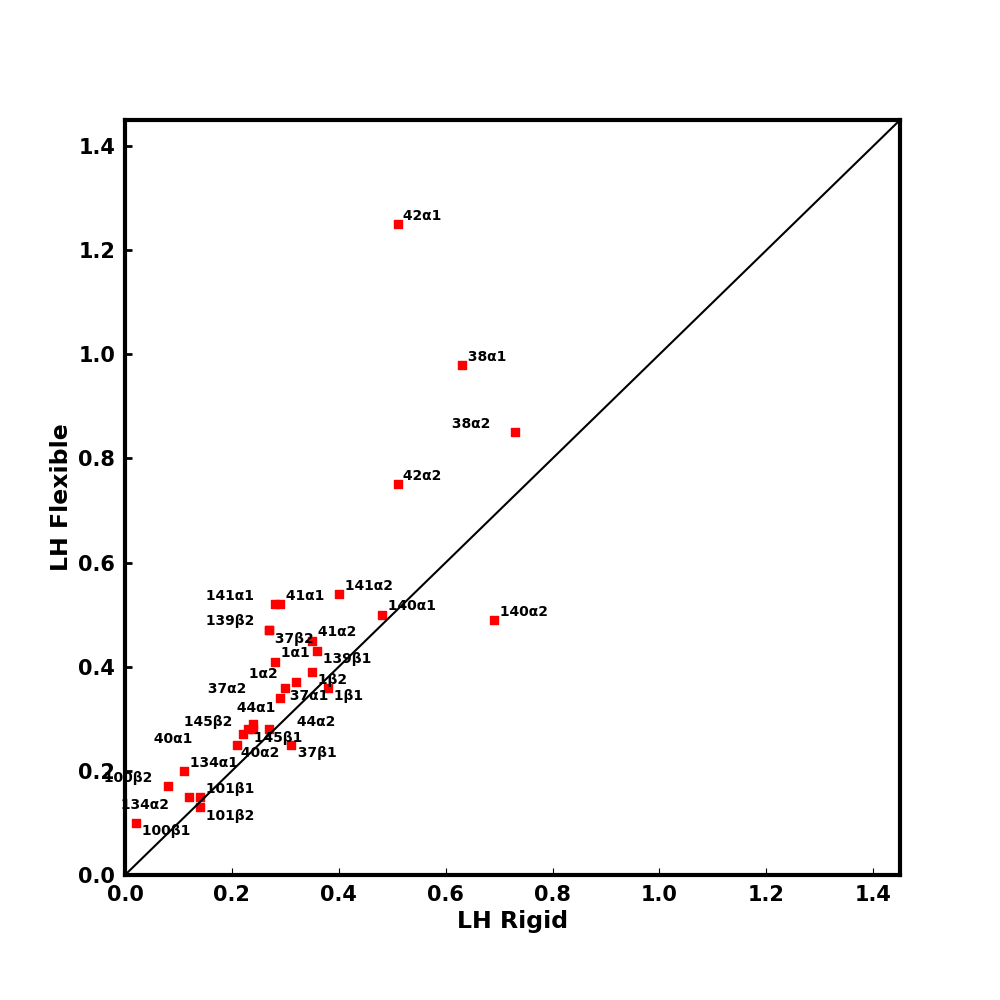}
\caption{Comparison of max$P({\rm LH})$ for rigid vs. flexible R$_4$ tetramer.}
\label{sifig:corr_rigid_flex_r4}
\end{figure}

\begin{figure}
  \centering
  \includegraphics[scale=0.5]{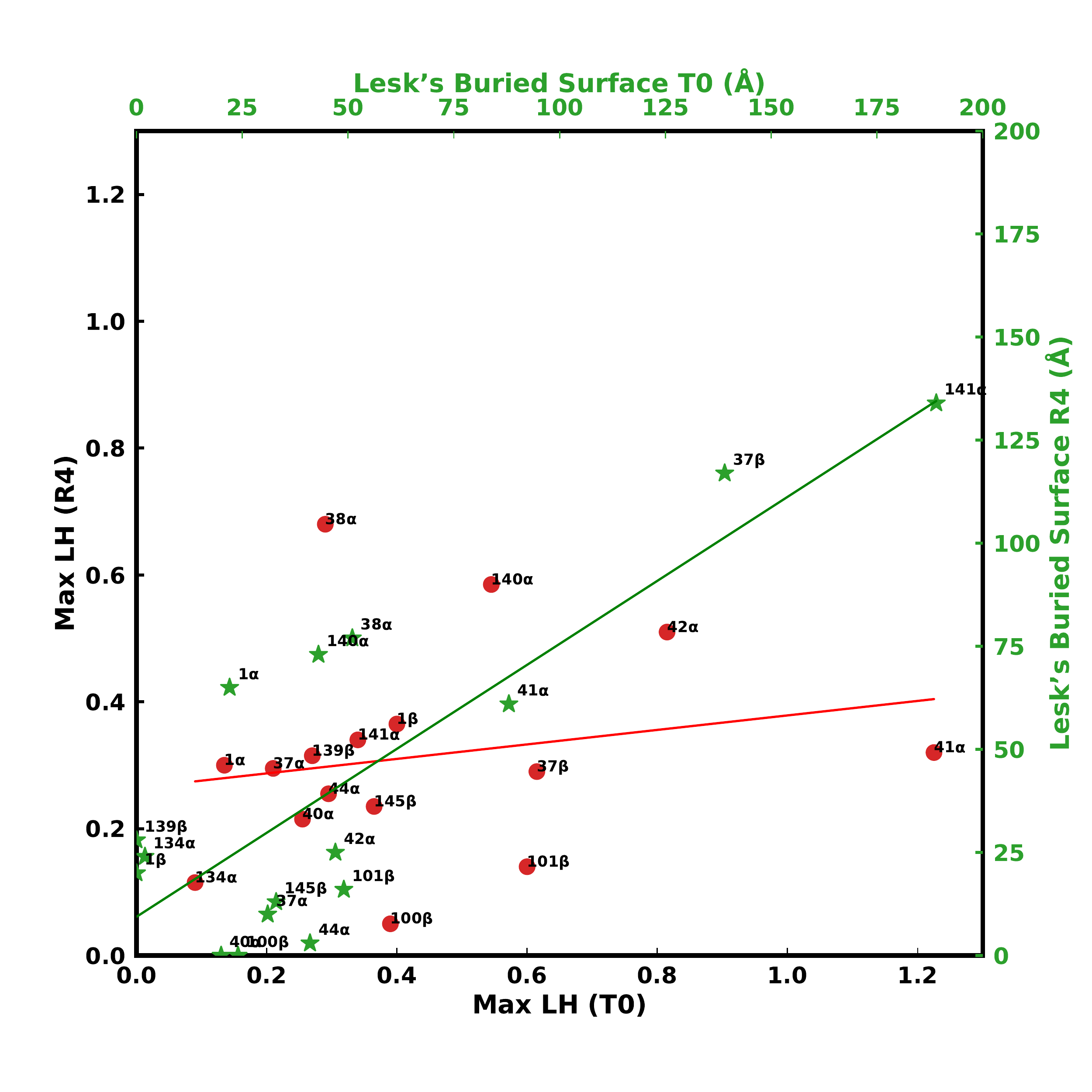}
\caption{Max$P({\rm LH})$ of T$_0$ vs R$_4$ for rigid tetramer. The
  max$P({\rm LH})$ is based on the average value between S1 and
  S2. The green asterisks represent Lesk's buried surface for T$_0$
  vs. R$_4$. The correlation coefficients between T$_0$ and R$_4$ are
  0.20 for red circles and 0.81 for green asterisks. The correlation
  fit line is also shown as red and green lines.}
\label{fig:lhavg.vs.lesk}
\end{figure}

\begin{figure}
  \centering \includegraphics[scale=0.5,
    angle=0]{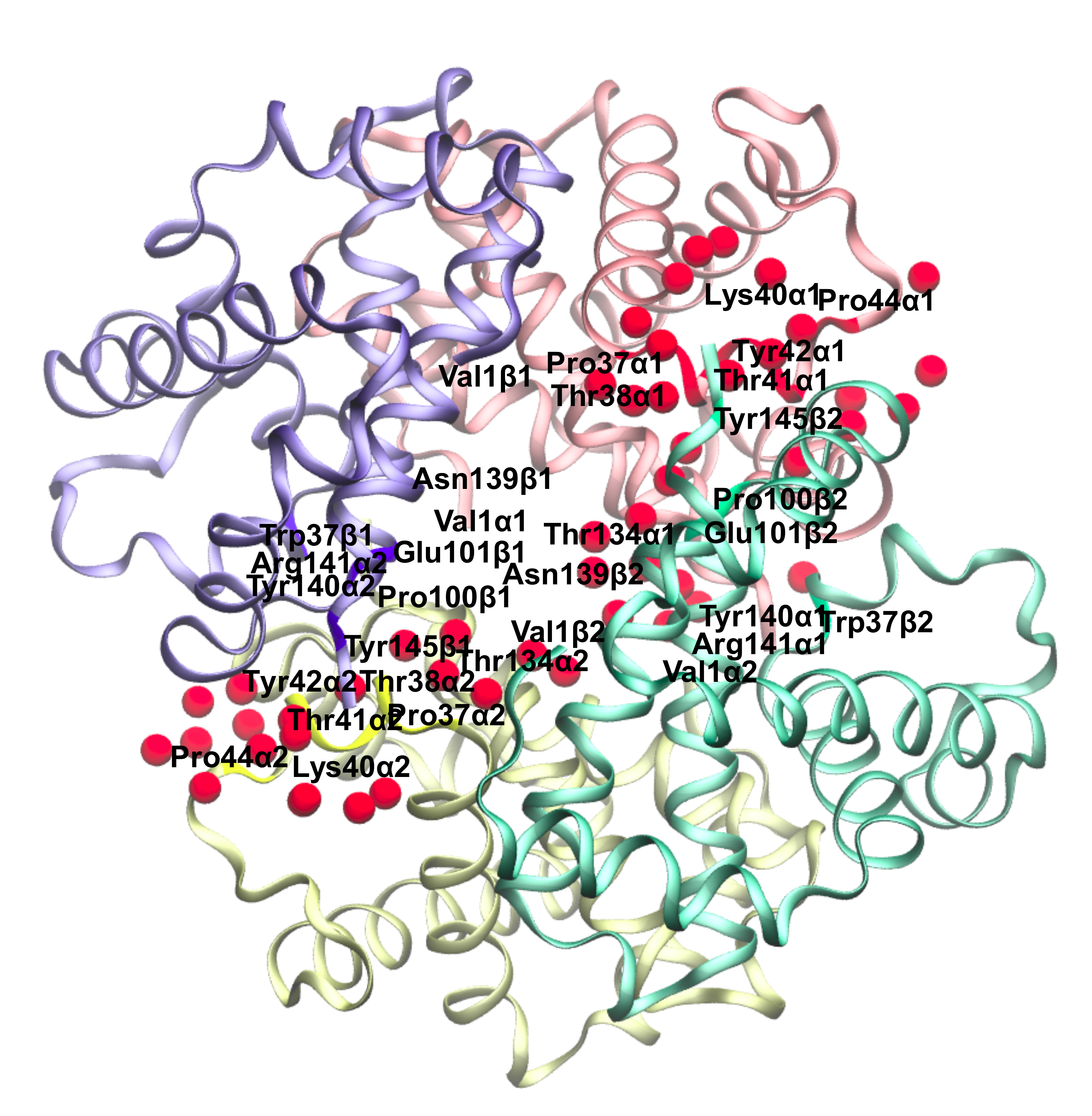}
\caption{For T$_0$, the water molecules (red spheres) within 3 \AA\/
  of any residue identified by the ticks in Table \ref{tab:tabres} as
  being at the $\alpha_1 , \beta_2$/$\alpha_2 , \beta_1$ interface
  with relevant residues labelled. The blue and green secondary
  structures refer to S1 and S2 and the relevant residues are
  labelled.}
\label{sifig:t0water}
\end{figure}

\begin{figure}
\centering \includegraphics[scale=0.9, angle=-90]{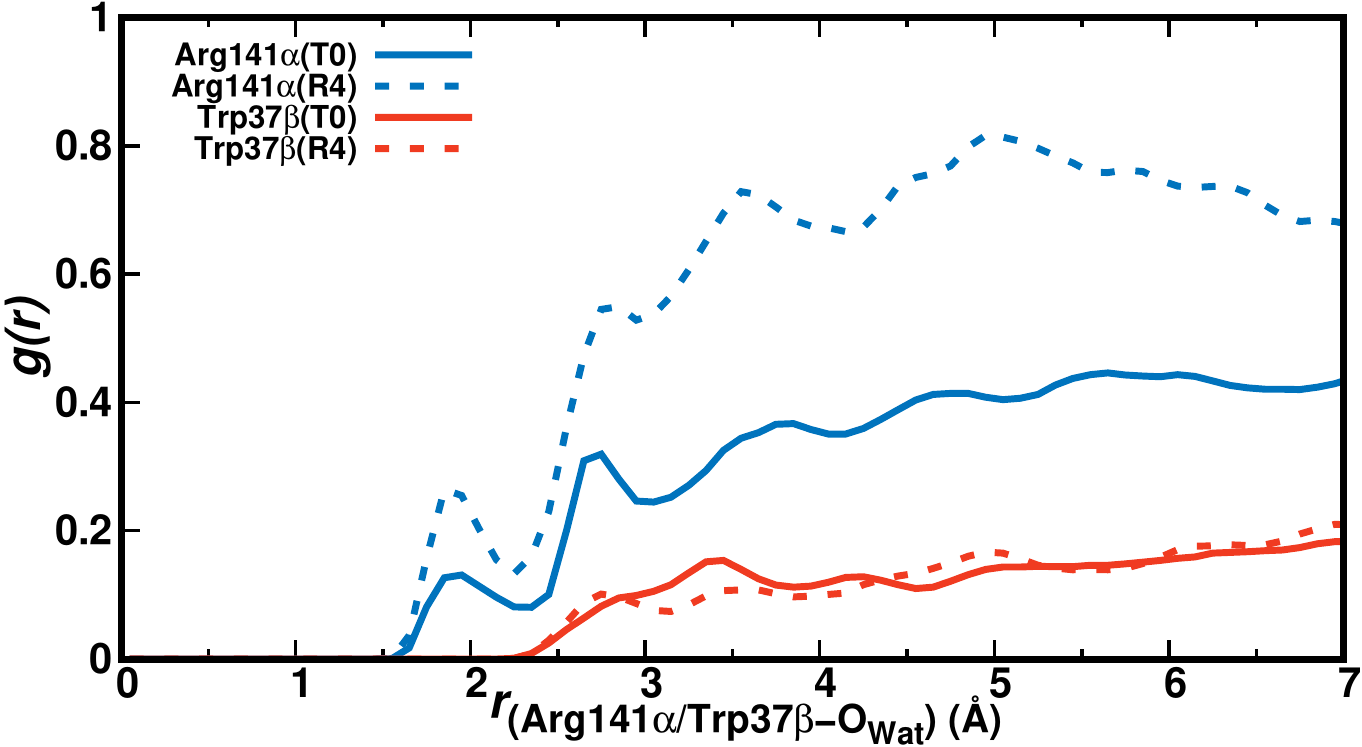}
\caption{$g(r)$ between solvent water and C$_\alpha$ for residues
  Arg141$\alpha$ and Trp37$\beta$ in T$_0$ and R$_4$ rigid tetramer.}
\label{sifig:gr}
\end{figure}

\begin{figure}
\centering
\includegraphics[scale=0.9,angle=-90]{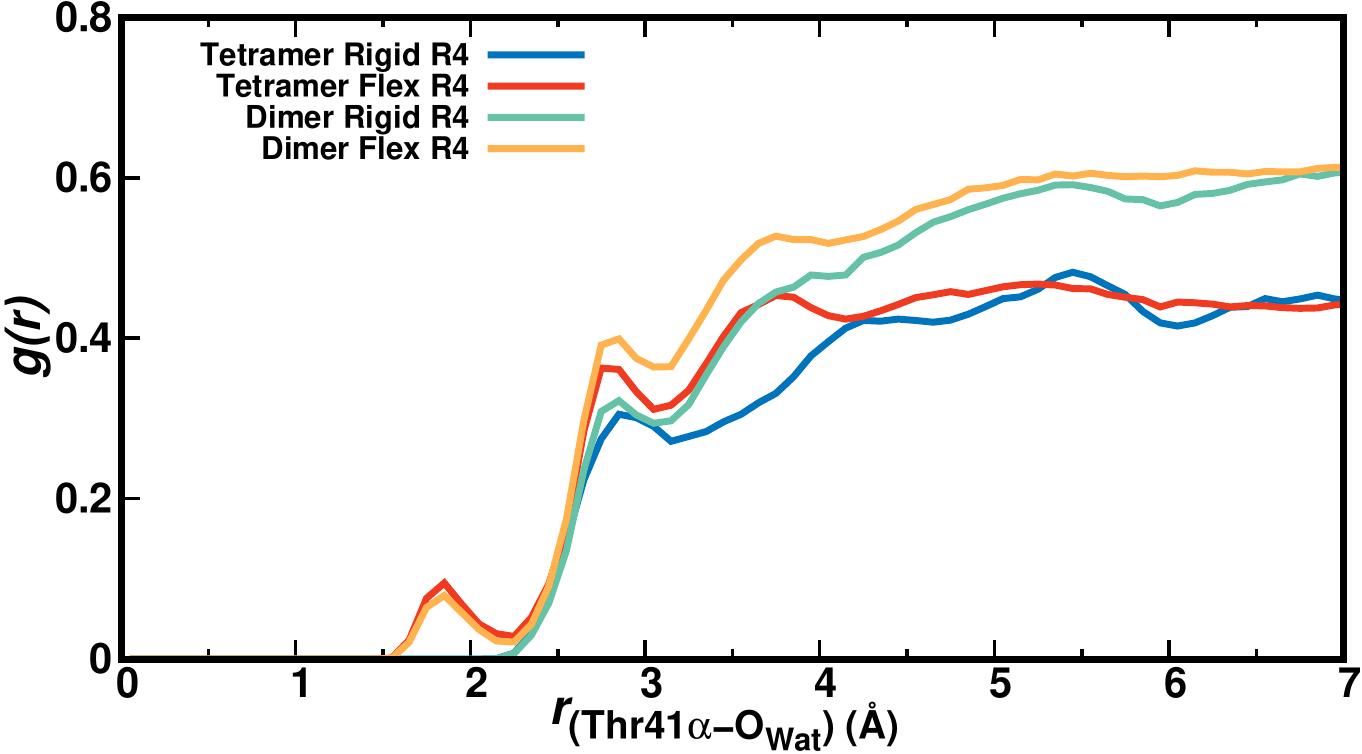}
\caption{$g(r)$ between solvent water and Thr41$\alpha$ for R$_4$
  rigid and flexible in dimer and tetramer.}
\label{sifig:gr3}
\end{figure}

\begin{figure}
\centering \includegraphics[scale=0.9,angle=-90]{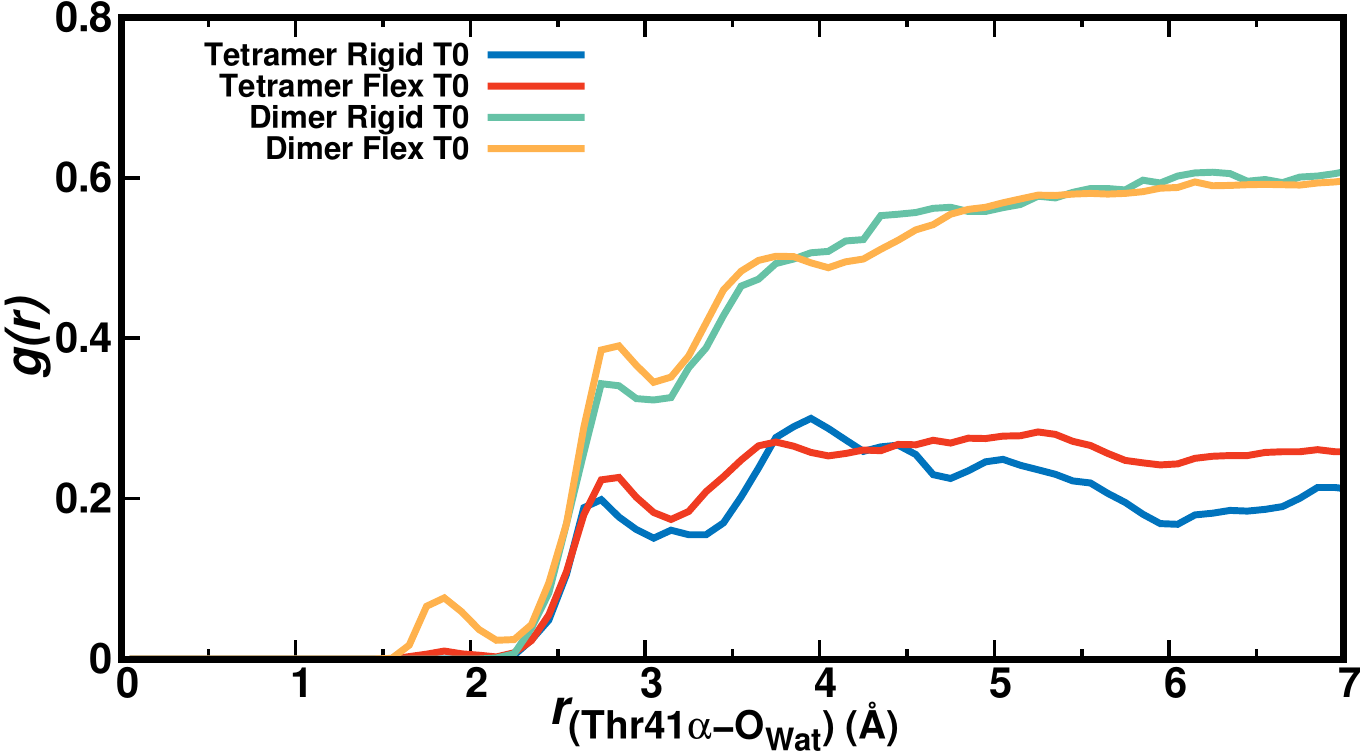}
\caption{$g(r)$ between solvent water and  Thr41$\alpha$
  for T$_0$ rigid and flexible in dimer and tetramer.}
\label{sifig:gr2}
\end{figure}

\begin{figure}
\centering \includegraphics[scale=1.2,angle=0]{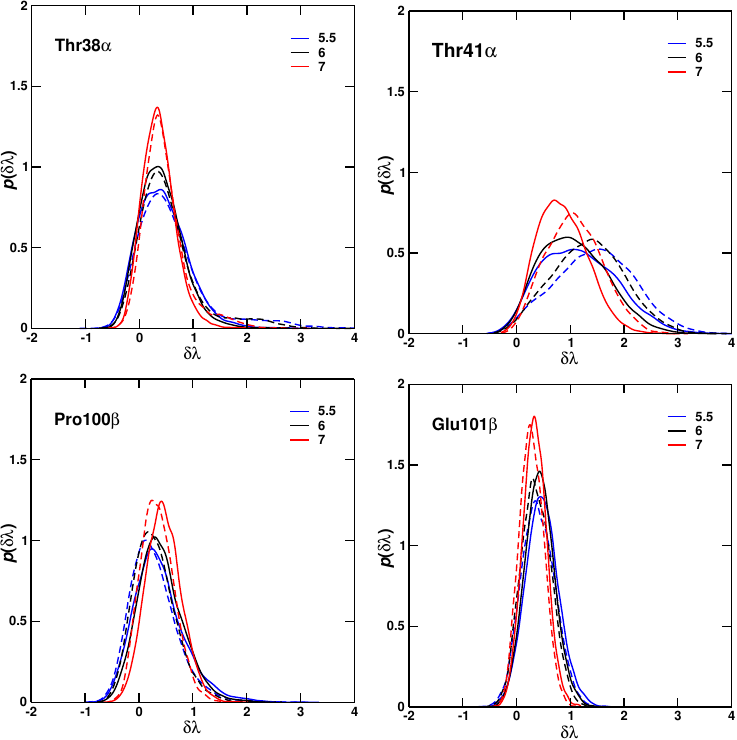}
\caption{$P({\rm LH})$ for 4 residues as indicated for the flexible
  T$_0$ tetramer for different cutoff distances (5.5 \AA\/ (blue), 6.0
  \AA\/ (black), and 7 \AA\/ (red)) considered in the LH
  analysis. $P({\rm LH})$ for subunits S1 and S2 are in solid and
  dashed lines, respectively.}
\label{sifig:plh}
\end{figure}

\end{document}